\documentclass[journal,twoside,web]{IEEEtran}
\usepackage{cite}
\usepackage{graphicx}
\usepackage{textcomp}
\usepackage{xr}
\usepackage{xcolor}
\usepackage{amssymb}
\usepackage{subfigure}
\usepackage{amsmath}
\usepackage{amsfonts}
\usepackage{multirow}
\usepackage{enumerate}
\usepackage{mathtools}
\usepackage{algorithm}
\usepackage[hidelinks]{hyperref}
\usepackage[noend]{algpseudocode}

\DeclareMathAlphabet{\mathpzc}{OT1}{pzc}{m}{it}

\newtheorem{theorem}{Theorem}

\begin{document}
\bibliographystyle{IEEEtran}
\title{Optimal Transport Driven Asymmetric Image-to-Image Translation for 
	Nuclei Segmentation of Histological Images}
\author{Suman Mahapatra and Pradipta Maji}

\maketitle

\begin{abstract}
Segmentation of nuclei regions from histological images enables morphometric 
analysis of nuclei structures, which in turn helps in the detection and 
diagnosis of diseases under consideration. To develop a nuclei segmentation 
algorithm, applicable to different types of target domain representations, 
image-to-image translation networks can be considered as they are invariant 
to target domain image representations. One of the 
important issues with image-to-image translation models is that they fail 
miserably when the information content between two image domains 
are asymmetric in nature. In this regard, the paper introduces a new deep 
generative model for segmenting nuclei structures from histological images. 
The proposed model considers an embedding space for handling 
information-disparity between information-rich histological image space and 
information-poor segmentation map domain. Integrating judiciously the 
concepts of optimal transport and measure theory, the model develops an 
invertible generator, which provides an efficient optimization framework 
with lower network complexity. The concept of invertible generator 
automatically eliminates the need of any explicit cycle-consistency loss. 
The proposed model also introduces a spatially-constrained squeeze 
operation within the framework of invertible generator to maintain spatial 
continuity within the image patches. The model provides a better trade-off 
between network complexity and model performance compared to other existing 
models having complex network architectures. 
An important finding is that the proposed model performs significantly better 
than the state-of-the-art methods in 87.5\% cases with respect to standard 
segmentation evaluation metrics, considering publicly available TCGA and CoNIC 
data sets. The performance of the proposed 
deep generative model, along with a comparison with state-of-the-art nuclei 
segmentation methods, is demonstrated on publicly available histological 
image data sets.
\end{abstract}


\section{Introduction}
\label{sec:introduction} 
\IEEEPARstart{M}{icroscopic} analysis of histological tissue images plays a very 
crucial role in the diagnosis, prognosis and treatment of cancer due to the 
abundance of phenotypic information present in histological images 
\cite{Elmore2015}. With the advent of digital pathology, computer-aided analysis 
of tissue images has become an integral step in the field of medical image 
analysis as it can ease pathologists' tasks.
The grades of different types of cancers are determined by analyzing different 
shapes and spatial arrangements of nuclei regions within a tissue specimen 
\cite{Filipczuk2013}. 
Hence, segmentation of nuclei structures from histological images has an 
immense importance in digital pathology as it provides significant 
morphological information, which helps in the therapeutic diagnosis of 
diseases under consideration. Segmenting nuclei structures from histological 
images is a challenging task as nuclei can exhibit different morphologies, 
color, texture or can be occluded partially by other nuclei or different 
cellular components. 

\par Nuclei segmentation is a process that focuses on the 
labeling of all pixels corresponding to each individual nucleus, and separating 
each nucleus from the background, other nuclei and different cellular components 
in a tissue image. 
Some earlier works in this domain have relied on energy-based 
methods, particularly the watershed algorithm. 
In \cite{Veta2013}, fast radial symmetry transform was employed to extract marker 
corresponding to each histological image and a sequence of different morphological 
operations was utilized to obtain the energy landscape. 
A variant of the watershed algorithm, which solely depends on the extracted energy 
landscape, was proposed in \cite{Ali2012}. It uses a combination of active contour 
and shape prior to extract nuclei regions from histological images. In 
\cite{Kwak2015}, a nuclei segmentation method was proposed, which utilizes the 
geometry of nuclei structures to compute the concavity of nuclei clusters. 
Based on the combination of double thresholding method and different 
morphological operations, a cell/nuclei segmentation approach was proposed in 
\cite{Chang2018}. 
However, these methods are sensitive to the choice of pre-defined 
parameters and assume the shapes of the nuclei apriori, leading to the 
extraction of faulty segmentation maps since nuclei can exhibit diverse shapes 
across different organs. 
Some shallow learning based models \cite{Song2018}, \cite{Song2019}, which 
perform convolution operation directly on histological images, utilize 
their abstract feature representations to extract pixel-wise labels from each 
input histological image. 
The main concern with these shallow learning based approaches is 
that they do not consider spatial dependency between image pixels and 
eventually extract inaccurate nuclei segmentation maps. 

\par Recent years have witnessed a huge surge in the application of deep 
learning based models in various medical image analysis tasks. In 
\cite{Ronneberger2015}, Ronneberger \textit{et~al.} proposed U-Net model, 
which has been applied in numerous biomedical image segmentation tasks. 
The U-Net architecture incorporates the concept of skip connections in 
traditional encoder-decoder framework to capture low-level semantic 
information. 
The U-Net++ model \cite{Zhou2018}, an extension of the U-Net architecture, 
enables feature propagation through densely interconnected skip connections.
Another extension of the U-Net architecture was proposed in \cite{Raza2019} 
for proper detection and segmentation of nuclei regions of different sizes. 
In a recent method \cite{Preity2024}, an attention mechanism 
based U-Net model has been introduced for the segmentation of blood vessels 
from retinal images. 
However, U-Net based methods often fail to detect clustered nuclei and 
are extremely sensitive to pre-defined parameters associated with weighted 
loss function. 
A supervised two-stage object segmentation method, called Mask-R-CNN, was 
proposed in \cite{He2017} for segmenting objects from natural as well as 
medical images. The Mask-R-CNN model predicts bounding boxes corresponding 
to each individual nucleus, and then, the nuclei regions are segmented 
inside the predicted bounding boxes. 
In \cite{Graham2019}, a unified fully convolutional network based model, 
named HoVer-Net, was proposed for simultaneous segmentation and classification 
of nuclei structures. This method utilizes the concepts of vertical and 
horizontal distance maps in order to separate occluded or touching nuclei 
structures. Most of these aforementioned approaches 
need a large number of annotated images, which are sometimes practically 
infeasible to have due to the time, effort and expert knowledge needed to 
annotate the data. 
A nuclei instance segmentation and classification approach, named Stardist, 
was proposed in \cite{Weigert2022}. Though Stardist performs well in case of 
convex object shapes, it fails miserably to detect heterogeneous nuclei shapes. 
This is due to the fact that Stardist assumes the nuclei shapes to be 
star-convex in nature, which is an impractical assumption about nuclei 
structures belonging to different organs of the body. 
Recently, Swin transformer, which is a type of vision transformer, 
has emerged as a new tool for image segmentation problems. In 
\cite{Lijin2024}, a Swin transformer based dual encoder-decoder architecture has 
been proposed for polyp segmentation from colonoscopy images.
A Swin transformer based multiple instance learning method, 
named Swin-MIL, has been proposed in \cite{Qian2022} for predicting masks from 
histological images. 
The main disadvantage with the vision transformer based models 
is that they are computationally expensive due to the large number of parameters 
involved and also need a substantial number of samples for training the models. 
In a recent method, called BoNuS \cite{Lin2024}, a binary mining loss term has 
been introduced to simultaneously learn nuclei interior and extract boundary 
information in order to segment nuclei structures from histological images.

\par Image-to-image (I2I) translation models aim to learn mappings between image 
domains while preserving certain high-level semantic information. 
One important characteristic of the I2I models is that they are invariant to 
different types of target domain image representations, shapes and sizes of the 
objects, which enhance the applicability of the translation 
models. 
To achieve unpaired training between two image domains, CycleGAN model was 
proposed in \cite{Zhu2017}, based on the concept of generative adversarial networks 
(GAN). Along with two generators: one for the forward 
mapping and the other one for inverse mapping corresponding to two image domains, 
CycleGAN model also introduced the concept 
of cycle-consistency loss term.  
In \cite{Mahmood2020}, conditional GAN with spectral normalization is employed 
in CycleGAN \cite{Zhu2017} framework for generating synthetic 
data and subsequently segmenting nuclei regions from histological images. 
A pathology-constrained CycleGAN model was proposed in \cite{Liu2021} 
for stain transfer, which 
can be utilized to address the nuclei segmentation problem. 
However, the CycleGAN based models fail in accurate segmentation of nuclei 
structures as the information content within the two image domains are asymmetric in 
nature. The histological images contain information regarding cell nuclei along with 
other tissue-level details, compared to segmentation maps, which contain information 
regarding only nuclei structures. 
Another concern with the CycleGAN based models is that they incorporate one generator 
and one discriminator corresponding to each of the image domains, which eventually 
increase the network complexity. 
In cycle-free CycleGAN model \cite{Kwon2021}, the concept of optimal transport is 
utilized to achieve unpaired I2I translation with lower network 
complexity. 
But, cycle-free CycleGAN model does not consider information disparity between 
histological image domain and segmentation map space. As a result, the extracted 
nuclei segmentation maps are expected to be less accurate. 
Hence, an I2I translation model needs to be developed to solve the 
nuclei segmentation problem by handling the information disparity present among two 
asymmetric image domains.

\par In this respect, the paper introduces a new deep generative model, 
termed as Ostrich 
(\underline{O}ptimal Tran\underline{s}por\underline{t} D\underline{r}iven 
Asymmetr\underline{ic} Image-to-Image Translation approac\underline{h}), 
for segmentation of nuclei structures from histological images. 
Particularly, the paper makes the following contributions:
\begin{itemize}
	\item The proposed Ostrich model is capable of handling information-asymmetry 
	between information-rich histological image space and information-poor 
	segmentation map domain by considering an embedding space to balance the 
	information disparity between two asymmetric image domains. 
	\item The proposed model combines 
	the concept of optimal transport with measure theory to develop an invertible 
	generator, which provides an efficient optimization framework with lower 
	network complexity as well as eliminates the need of any explicit 
	cycle-consistency loss term.
	\item The model also introduces a 
	simple yet effective spatially-constrained squeeze operation within the 
	framework of coupling layers associated with invertible generator to maintain 
	intra-image spatial continuity.	
	\item The proposed model provides a better trade-off 
	between network complexity and model performance compared to other existing 
	models having more complex network architectures. 
\end{itemize}
The efficacy of the proposed Ostrich model, along with a 
comparison with state-of-the-art nuclei segmentation approaches, has been 
established on publicly available H\&E stained histological image data sets.

\section{Ostrich: Proposed Method}
\label{proposed_method}
In this section, a deep generative model, based on the theory of optimal transport 
\cite{Villani2008}, is presented for nuclei segmentation from histological 
images. The proposed model is inspired by the cycle-free CycleGAN model introduced 
in \cite{Kwon2021}. 
The main difference between the proposed Ostrich model and 
\cite{Kwon2021} is that the Ostrich model tries to perform image-to-image translation 
between two image domains with information imbalance, which the 
cycle-free CycleGAN model \cite{Kwon2021} is not capable of dealing with. 


\subsection{Problem Statement}
\label{problem_statement}
Given a histological image domain $\mathcal{X}$, containing a set of $n$ number of images 
$\{x_{i}: i = 1, 2, \cdots, n\}$, the aim is to develop a model, which takes each 
histological image $x_{i} \in \mathcal{X} \subset \mathbb{R}^{H\times W\times 3}$ 
as input and outputs corresponding nuclei segmentation map $y_{i}$. The set of 
these output maps $\{y_{i}\}$ forms segmentation map domain $\mathcal{Y}$, that is, 
$y_{i} \in \mathcal{Y} \subset \mathbb{R}^{H\times W}$. 
As described in the previous section, there is an observable 
information-imbalance between the two image domains: histological image space $\mathcal{X}$ and 
segmentation map domain $\mathcal{Y}$. Hence, the goal of the current study is to design a 
deep generative model that, with proper training, will be able to generate 
segmentation annotation map corresponding to each input histological image by 
handling the information-asymmetry between two image domains.

\subsection{Cycle-free Asymmetric I2I Translation}
\label{AsymmGAN_derivation}
This subsection introduces an invertible I2I translation 
approach for solving the problem of nuclei segmentation of histological images. 
The following analysis establishes that two generators and one discriminator 
are sufficient to solve this problem.

\par To balance the information-asymmetry while translating from information-rich 
image space $\mathcal{X}$ to information-poor segmentation map domain $\mathcal{Y}$, an 
embedding space $\mathcal{Z}$ is considered in this study. 
Let, the histological image space $\mathcal{X}$ be equipped with probability 
measure $\mu$, whereas the annotation image space $\mathcal{Y}$ and the embedding 
space $\mathcal{Z}$ be 
equipped with probability measures 
$\nu$ and $\eta$, respectively.
The aim of the proposed model is to transport the joint 
distribution of $\nu$ and $\eta$ of the annotation domain $\mathcal{Y}$ and 
embedding space $\mathcal{Z}$, respectively, that is, $(\nu \times \eta)$, 
to distribution $\mu$ of the image space $\mathcal{X}$ so that the joint distribution 
can mimic the distribution of the histological image space.

\par In the proposed model, three generators $F_{\phi}$, 
$E_{\omega}$, and $G_{\theta}$ are considered. The forward operator $F_{\phi}$ 
``pushes forward'' measure 
$\mu$ in image space $\mathcal{X}$ to measure $\nu_{\phi}$ in the annotation space 
$\mathcal{Y}$, that is, $F_{\phi}: \mathcal{X} \to \mathcal{Y}$. Similarly, the 
transportation from probability space $(\mathcal{X}, \mu)$ to $(\mathcal{Z}, \eta)$ is 
performed by the forward operator $E_{\omega}$, which pushes forward measure $\mu$ in 
image space $\mathcal{X}$ to $\eta_{\omega}$ in the embedding space $\mathcal{Z}$, that 
is, $E_{\omega}: \mathcal{X} \to \mathcal{Z}$. On the other hand, the generator $G_{\theta}$ 
transports the product measure $(\nu \times \eta)$ in the product space 
$(\mathcal{Y} \times \mathcal{Z})$ to $\mu_{\theta}$ in the target histological image 
space $\mathcal{X}$, that is, 
$G_{\theta}: (\mathcal{Y} \times \mathcal{Z}) \to \mathcal{X}$. Hence, the optimal 
transport problem in this study can be solved by minimizing the distances 
$d(\mu, \mu_{\theta})$, $d(\nu, \nu_{\phi})$, and $d(\eta, \eta_{\omega})$, where 
$d(\alpha, \beta)$ denotes statistical distance between two probability measures 
$\alpha$ and $\beta$. So, to solve this simultaneous statistical distance minimization 
problem, the following optimal transport problem needs to be solved:
\begin{align}
\label{OT_Primal_Cost}
\inf_{\pi \in \Pi(\mu, \nu, \eta)} 
\int \limits_{\mathcal{X} \times \mathcal{Y} \times \mathcal{Z}} 
\mathpzc{C}(x, y, z; G_{\theta}, F_{\phi}, E_{\omega}) d\pi(x, y, z)
\end{align}   
where $\Pi(\mu, \nu, \eta)$ denotes the set of all joint probability distributions 
with the marginals $\mu$, $\nu$, and $\eta$. 
In this study, the transportation cost is defined as follows:
\begin{align}
\label{OT_Primal_Terms}
&\mathpzc{C}(x, y, z; G_{\theta}, F_{\phi}, E_{\omega}) = 
\lVert x - G_{\theta}(y, z) \rVert + \nonumber \\
&\quad \quad \quad \quad \lVert F_{\phi}(x) - y \rVert + 
\lVert E_{\omega}(x) - z \rVert .
\end{align}
\par At the initial stage, the output segmentation maps, generated by the forward 
operator $F_{\phi}$, are far from the real data distribution in $\mathcal{Y}$. So, 
to push the generation in the proper direction, a regularization term 
is introduced in the current study, which is defined as follows: 
\begin{align}
\label{regularization_term}
&l_{ssim}(F_{\phi}) = \lVert 1 - SSIM(F_{\phi}(x), VO(f_{rgb}^{hed}(x)[:, :, 0]))\rVert.
\end{align}
Here, $SSIM$ denotes a structural similarity measurement index, called 
structural similarity index measure \cite{Zhou2004}, 
$f_{rgb}^{hed}(.)$ represents a function for converting a color image from RGB 
to Hematoxylin-Eosin-DAB (HED) colorspace. As nuclei regions in a H\&E-stained 
image are generally 
highlighted by hematoxylin (H) stain, the H-channel of an image $x \in \mathcal{X}$ is 
considered through $f_{rgb}^{hed}(x)[:, :, 0]$. $VO(.)$ denotes a 
transformation, which takes the H-channel image as input and outputs corresponding 
segmentation map by leveraging Voronoi labeling on thresholded H-channel image 
due to Otsu \cite{Otsu1979}. 
The aim is to maximize the structural similarity 
between the generated output and the transformed output through this regularization 
term. So, considering the regularization term, the overall transportation cost is 
defined as follows: 
\begin{align}
\label{OT_Primal_with_SSIM}
&\mathpzc{C}(x, y, z; G_{\theta}, F_{\phi}, E_{\omega}) = 
\lVert x - G_{\theta}(y, z) \rVert + 
\lVert F_{\phi}(x) - y \rVert + \nonumber \\ 
&\quad \quad \quad \quad \quad \quad \quad \quad  
\lVert E_{\omega}(x) - z \rVert + \lambda_{1} 
l_{ssim}(F_{\phi})
\end{align}
where $\lambda_{1}$ denotes the relative importance of the regularization term 
$l_{ssim}(F_{\phi})$, defined in (\ref{regularization_term}). 
The value of $\lambda_{1}$ decreases gradually during the training.
The computation of the regularization term is trivial, but from the perspective of 
optimal transport, the computation of the first three terms requires corresponding 
dual formulation \cite{Villani2008}, \cite{Peyre2019}.

\par The conversion from primal to dual form is provided in the 
appendix. 
The dual formulation corresponding to the optimal transport problem presented 
in (\ref{OT_Primal_Cost}), with the transportation cost 
$\mathpzc{C}(x, y, z; G_{\theta}, F_{\phi}, E_{\omega})$
, defined in (\ref{OT_Primal_with_SSIM}), 
can be represented as follows:
\begin{align}
\label{OT_Dual}
\min \limits_{\theta, \phi, \omega} ~ \max \limits_{\varphi, \psi, \xi}
l(G_{\theta}, F_{\phi}, E_{\omega}; \varphi, \psi, \xi)
\end{align}
where $\varphi$, $\psi$ and $\xi$ represent three discriminators corresponding 
to three domains $\mathcal{X}$, $\mathcal{Y}$ and $\mathcal{Z}$, respectively. 
Here, the loss $l(G_{\theta}, F_{\phi}, E_{\omega}; \varphi, \psi, \xi)$ is 
defined as follows:
\begin{align}
\label{OT_Dual_total}
&l(G_{\theta}, F_{\phi}, E_{\omega}; \varphi, \psi, \xi) = 
l_{GAN}(G_{\theta}, F_{\phi}, E_{\omega}; \varphi, \psi, \xi) \nonumber \\
&\quad \quad \quad \quad 
+\lambda_{2} l_{cycle}(G_{\theta}, F_{\phi}, E_{\omega}) + \lambda_{1} l_{ssim}(F_{\phi})
\end{align}
where the optimal transport based GAN loss term 
$l_{GAN}(G_{\theta}, F_{\phi}, E_{\omega}; \varphi, \psi, \xi)$ is defined as 
follows:
\begin{align}
\label{OT_GAN_loss}
&\quad \quad \quad \quad \quad \quad 
l_{GAN}(G_{\theta}, F_{\phi}, E_{\omega}; \varphi, \psi, \xi) 
= \nonumber \\ 
&\max \limits_{\varphi \in Lip_{1}(\mathcal{X})} 
\left[\int \limits_{\mathcal{X}} \varphi(x) d\mu(x) - 
\int \limits_{\mathcal{Y} \times \mathcal{Z}} \varphi(G_{\theta}(y,z)) 
d(\nu \times \eta)(y,z)\right] \nonumber \\
&\quad \quad + \max \limits_{\psi \in Lip_{1}(\mathcal{Y})} 
\left[\int \limits_{\mathcal{Y}} \psi(y) d\nu(y) - 
\int \limits_{\mathcal{X}} \psi(F_{\phi}(x)) d\mu(x)\right] \nonumber \\
&\quad \quad + \max \limits_{\xi \in Lip_{1}(\mathcal{Z})} 
\left[\int \limits_{\mathcal{Z}} \xi(z) d\eta(z) - 
\int \limits_{\mathcal{X}} \xi(E_{\omega}(x)) d\mu(x)\right]
\end{align}
where $Lip_{m}(D)$ represents the space of all $m$-Lipschitz functions defined over 
domain $D$. $G_{\theta}$ is a $(\nu \times \eta)$-measurable 
function defined on the product space $(\mathcal{Y}\times \mathcal{Z})$. So, the sections 
of $G_{\theta}$, that is, ${G_{\theta}}_{y}$ and ${G_{\theta}}^{z}$, are 
$\eta$-measurable function defined on $\mathcal{Z}$ space 
and $\nu$-measurable function defined on domain 
$\mathcal{Y}$, respectively \cite{Cohn2013}.

The parameter $\lambda_{2}$ in (\ref{OT_Dual_total}) denotes the relative 
importance of the cycle-consistency loss term 
$l_{cycle}(G_{\theta}, F_{\phi}, E_{\omega})$, 
which can be defined as follows:
\begin{align}
\label{loss_cycle-consistency}
&\quad \quad l_{cycle}(G_{\theta}, F_{\phi}, E_{\omega}) = 
\int \limits_{\mathcal{X}}\lVert x - {G_{\theta}}^{z}(F_{\phi}(x)) \rVert d\mu(x) \nonumber \\ 
&+ \int \limits_{\mathcal{X}}\lVert x - {G_{\theta}}_{y}(E_{\omega}(x)) \rVert d\mu(x) 
+\int \limits_{\mathcal{Y}}\lVert y - F_{\phi}({G_{\theta}}^{z}(y)) \rVert d\nu(y) \nonumber \\
& \quad \quad \quad \quad \quad \quad + \int \limits_{\mathcal{Z}}\lVert z - E_{\omega}({G_{\theta}}_{y}(z)) \rVert d\eta(z)
\end{align}


\begin{theorem}
	\label{propose}
	If the sections of generator $G_{\theta}$, that is, ${G_{\theta}}_{y}$ and 
	${G_{\theta}}^{z}$, are considered to be the inverse operators of forward 
	operators $E_{\omega}$ and $F_{\phi}$, respectively, that is, 
	${G_{\theta}}_{y} = {E_{\omega}}^{-1}$ and ${G_{\theta}}^{z} = {F_{\phi}}^{-1}$ 
	and ${G_{\theta}}^{z}$ is a $p$-Lipschitz function in domain $\mathcal{Y}$, 
	then the problem in (\ref{OT_Dual}) with $p=1$, can be 
	represented by the following equivalent problem:
	\begin{align}
	\min \limits_{\phi, \omega} \max \limits_{\psi, \xi} l(F_{\phi}, E_{\omega}; \psi, \xi)
	\end{align}
	\begin{align}
	&\text{where} ~~~ l(F_{\phi}, E_{\omega}; \psi, \xi) =2 l_{GAN}(F_{\phi}; \psi) \nonumber \\ 
	&\quad \quad \quad \quad + l_{GAN}(E_{\omega}; \xi)  + \lambda_{1} l_{ssim}(F_{\phi})
	\end{align}
	and
	\begin{align}
	l_{GAN}(F_{\phi}; \psi) 
	= \max \limits_{\psi \in Lip_{1}(\mathcal{Y})} 
	\int \limits_{\mathcal{Y}} \psi(y) d\nu(y) - 
	\int \limits_{\mathcal{X}} \psi(F_{\phi}(x)) d\mu(x) ,
	\end{align}
	\begin{align}
	l_{GAN}(E_{\omega}; \xi) 
	= \max \limits_{\xi \in Lip_{1}(\mathcal{Z})} 
	\int \limits_{\mathcal{Z}} \xi(z) d\eta(z) - 
	\int \limits_{\mathcal{X}} \xi(E_{\omega}(x)) d\mu(x) .
	\end{align}
\end{theorem}

\par \textit{Proof.} The invertibility conditions ${G_{\theta}}_{y} = 
{E_{\omega}}^{-1}$ and ${G_{\theta}}^{z} = {F_{\phi}}^{-1}$ ensure that 
${G_{\theta}}_{y}(E_{\omega}(x)) = x$, ${G_{\theta}}^{z}(F_{\phi}(x)) = x$, 
$F_{\phi}({G_{\theta}}^{z}(y)) = y$ and $E_{\omega}({G_{\theta}}_{y}(z)) = z$. So, 
it is evident that if the above invertibility conditions hold, then the 
cycle-consistency loss term $l_{cycle}(G_{\theta}, F_{\phi}, E_{\omega})$ vanishes.  
\par Let, $\varphi^{*} \in Lip_{1}(\mathcal{X})$ be the optimum $1$-Lipschitz function 
in domain $\mathcal{X}$ for which the first sub-expression from (\ref{OT_GAN_loss}) 
attains maximum value. So, 
\begin{align}
\label{GAN_x_derive}
&\max \limits_{\varphi \in Lip_{1}(\mathcal{X})} 
\left[\int \limits_{\mathcal{X}} \varphi(x) d\mu(x) - 
\int \limits_{\mathcal{Y} \times \mathcal{Z}} \varphi(G_{\theta}(y,z)) 
d(\nu \times \eta)(y,z)\right] \nonumber \\
& = \int \limits_{\mathcal{X}} \varphi^{*}(x) d\mu(x) - 
\int \limits_{\mathcal{Y} \times \mathcal{Z}} \varphi^{*}(G_{\theta}(y,z)) 
d(\nu \times \eta)(y,z) . 
\end{align}
Let, the two terms in (\ref{GAN_x_derive}) be denoted by $J_{1}$ and $J_{2}$, 
respectively, that is, 
\begin{align}
\label{GAN_x_derive_first}
J_{1} = \int \limits_{\mathcal{X}} {\varphi}^{*}(x) d\mu(x)
\end{align}

\begin{align}
\label{GAN_x_derive_second}
\text{and} ~~~ J_{2} = -\int \limits_{\mathcal{Y} \times \mathcal{Z}} 
{\varphi}^{*}(G_{\theta}(y,z)) d(\nu \times \eta)(y,z) .
\end{align}
As ${G_{\theta}}^{z}$ is designed to be the inverse of forward operator 
$F_{\phi}$, the following formulation can be derived from 
(\ref{GAN_x_derive_first}):
\begin{align}
\label{GAN_x_derive_first_contd2}
& J_{1} = \int \limits_{\mathcal{X}} 
{\varphi}^{*}({G_{\theta}}^{z}(F_{\phi}(x))) d\mu(x)
= \max \limits_{{\psi}_{\theta} \in \Phi_{1}} \int \limits_{\mathcal{X}} 
{\psi}_{\theta}(F_{\phi}(x)) d\mu(x)
\end{align}
where the set $\Phi_{1}$ is defined as follows:
\begin{align}
\label{Set_phi_1_define}
\Phi_{1} = \{\hat{\psi}\lvert \hat{\psi} = \varphi \circ {G_{\theta}}^{z}, 
\varphi \in Lip_{1}(\mathcal{X}), {G_{\theta}}^{z} \in Lip_{1}(Y)\}.
\end{align}
Here, `$\circ$' denotes the composition of functions.
Now, (\ref{GAN_x_derive_second}) can be rewritten as follows:
\begin{align}
\label{GAN_x_derive_second_compose}
& \quad J_{2} = \max \limits_{{\mathpzc{H}}_{\theta} \in \Phi_{2}}
\left [-\int \limits_{\mathcal{Y} \times \mathcal{Z}} {\mathpzc{H}}_{\theta}(y, z) 
d(\nu \times \eta)(y,z)\right].
\end{align}
Here, the set $\Phi_{2}$ is defined as follows:
\begin{align}
\label{Set_phi_2_define}
\Phi_{2} = \{\mathpzc{H}\lvert \mathpzc{H} = \varphi \circ G_{\theta}, 
\varphi \in Lip_{1}(\mathcal{X})\}.
\end{align}
Using Fubini's theorem on product measures \cite{Cohn2013}, 
(\ref{GAN_x_derive_second_compose}) can be formulated as follows:
\begin{align}
\label{GAN_x_derive_second_derive}
J_{2} = \max \limits_{{\mathpzc{H}}_{\theta} \in \Phi_{2}}
\left[-\int \limits_{\mathcal{Y}} \left\{\int \limits_\mathcal{Z} 
{\mathpzc{H}}_{\theta}(y, z) d\eta(z)\right\} d\nu(y)\right].
\end{align}
The expression $\int \limits_\mathcal{Z}{\mathpzc{H}}_{\theta}(y, z) d\eta(z)$ 
in (\ref{GAN_x_derive_second_derive}) is a $\nu$-measurable function and is defined 
on space $\mathcal{Y}$. Let, the expression be denoted by ${\mathpzc{M}}_{\theta}(y)$.  
For any two points $y_{1}, y_{2} \in \mathcal{Y}$, 
\begin{align}
\label{M_theta_lipschitz_property}
&\quad \quad \quad \quad \quad \quad \quad 
\lVert {\mathpzc{M}}_{\theta}(y_{1}) - {\mathpzc{M}}_{\theta}(y_{2}) \rVert \nonumber \\
&\quad \quad =\lVert \int \limits_\mathcal{Z}{\mathpzc{H}}_{\theta}(y_{1}, z) d\eta(z) - 
\int \limits_\mathcal{Z}{\mathpzc{H}}_{\theta}(y_{2}, z) d\eta(z)\rVert \nonumber \\
&\quad \quad \quad\le \int \limits_\mathcal{Z} \lVert {\mathpzc{H}}_{\theta}(y_{1}, z) 
- {\mathpzc{H}}_{\theta}(y_{2}, z)\rVert d\eta(z) \nonumber \\
&\quad \quad \quad \le \int \limits_\mathcal{Z} \lVert G_{\theta}(y_{1}, z) 
- G_{\theta}(y_{2}, z)\rVert d\eta(z) \nonumber \\
&\quad \quad \quad \quad \text{[as ${\mathpzc{H}}_{\theta} = 
	\varphi \circ G_{\theta}$ and $\varphi \in Lip_{1}(\mathcal{X})$]} \nonumber \\
& = \int \limits_\mathcal{Z} \lVert {G_{\theta}}^{z}(y_{1}) 
- {G_{\theta}}^{z}(y_{2})\rVert d\eta(z) 
\le \lVert y_{1} - y_{2} \rVert \int \limits_\mathcal{Z} d\eta(z) \nonumber \\
&\quad \quad \quad \quad = \lVert y_{1} - y_{2} \rVert 
~~~\text{[as ${G_{\theta}}^{z} \in Lip_{1}(\mathcal{Y})$]} .
\end{align}
Thus, (\ref{GAN_x_derive_second_derive}) can be reformulated as follows:
\begin{align}
\label{GAN_x_derive_second_reformulated}
J_{2} = \max \limits_{{\mathpzc{M}}_{\theta} \in \Phi_{3}} 
\left[-\int \limits_{\mathcal{Y}} {\mathpzc{M}}_{\theta}(y) d\nu(y) \right]
\end{align}
where the set $\Phi_{3}$ can be defined as follows:
\begin{align}
\label{Set_phi_3_define}
\Phi_{3} = \{\mathpzc{M}\lvert \mathpzc{M} = 
\int \limits_\mathcal{Z} \mathpzc{H}(y, z) d\eta(z), \mathpzc{H} \in \Phi_{2}\} .
\end{align}

Hence, the set $\Phi_{3}$ constructs 
a subset of the set of all $1$-Lipschitz functions on domain $\mathcal{Y}$. Again, 
for any element $\hat{\psi}_{1} \in \Phi_{1}$, $\hat{\psi}_{1}$ is of the form 
$\hat{\psi}_{1} = \varphi_{1} \circ \mathpzc{g}$, where 
$\varphi_{1} \in Lip_{1}(\mathcal{X})$ and $\mathpzc{g} \in Lip_{1}(\mathcal{Y})$. 
So, $\mathpzc{g}$ can potentially be any element from the set 
$\Phi_{3}$ and $\hat{\psi}_{1}$ can map the element to $Lip_{1}(\mathcal{Y})$ 
within or beyond $\Phi_{3}$. So, the above argument ensures that $\Phi_{3}$ is 
contained in $\Phi_{1}$, that is, $\Phi_{3} \subseteq \Phi_{1}$. Hence, by combining 
(\ref{GAN_x_derive_first_contd2}) and (\ref{GAN_x_derive_second_reformulated}), 
the following relation can be deduced:
\begin{align}
\label{J1_J2_combined}
&\quad J_{1} + J_{2} \le \max \limits_{\psi^{'} \in \Phi_{1}} 
\left[\int \limits_{\mathcal{X}} \psi^{'}(F_{\phi}(x)) d\mu(x) 
- \int \limits_{\mathcal{Y}} \psi^{'}(y) d\nu(y)\right] \nonumber \\
&\quad \quad \quad = \max \limits_{\tilde{\psi} \in \Phi_{1}} 
\left[\int \limits_{\mathcal{Y}} \tilde{\psi}(y) d\nu(y) - 
\int \limits_{\mathcal{X}} \tilde{\psi} (F_{\phi}(x)) d\mu(x) \right] \nonumber \\
&\text{[equality comes from the fact: $\tilde{\psi} = - \psi^{'}$ and thus 
	$\tilde{\psi} \in \Phi_{1}$]}.
\end{align} 
Now, by extending the function space from set $\Phi_{1}$ to $Lip_{1}(\mathcal{Y})$, 
the following relation can be obtained from (\ref{GAN_x_derive}) and 
(\ref{J1_J2_combined}):
\begin{align}
\label{relation_xyz_to_yx_domain}
&\max \limits_{\varphi \in Lip_{1}(\mathcal{X})} 
\left[\int \limits_{\mathcal{X}} \varphi(x) d\mu(x) - 
\int \limits_{\mathcal{Y} \times \mathcal{Z}} \varphi(G_{\theta}(y,z)) 
d(\nu \times \eta)(y,z)\right] \nonumber \\
&~~ \le \max \limits_{\psi \in Lip_{1}(\mathcal{Y})} 
\left[\int \limits_{\mathcal{Y}} \psi(y) d\nu(y) - 
\int \limits_{\mathcal{X}} \psi(F_{\phi}(x)) d\mu(x)\right].
\end{align}

Next, it must be shown that the upper bound in 
(\ref{relation_xyz_to_yx_domain}) is tight. 
Let, $\psi^{*}$ be the maximizer for 
(\ref{relation_xyz_to_yx_domain}). To show the tightness of the upper bound, it 
is important to show that there exists $\varphi \in Lip_{1}(\mathcal{X})$ 
such that, for $z \in \mathcal{Z}$:
\begin{align*}
\psi^{*}(y) = \varphi({G_{\theta}}^{z}(y)) = \varphi(G_{\theta}(y, z)), 
~~~ \forall y \in \mathcal{Y} .
\end{align*}  
Due to the invertibility condition ${G_{\theta}}^{z} = {F_{\phi}}^{-1}$, 
an element $x \in \mathcal{X}$ can always be found such that 
$x = {F_{\phi}}^{-1}(y) = {G_{\theta}}^{z}(y)$ for all $y \in \mathcal{Y}$. So, 
\begin{align*}
\varphi(x) = \varphi({F_{\phi}}^{-1}(F_{\phi}(x)))
= \varphi({G_{\theta}}^{z}(F_{\phi}(x))) = \psi^{*}(F_{\phi}(x)),
\end{align*}
which achieves the above upper bound. Hence, the following interesting relation can be 
obtained from (\ref{relation_xyz_to_yx_domain}):
\begin{align}
\label{relation_xyz_to_yx_domain_final}
&\max \limits_{\varphi \in Lip_{1}(\mathcal{X})} 
\left[\int \limits_{\mathcal{X}} \varphi(x) d\mu(x) - 
\int \limits_{\mathcal{Y} \times \mathcal{Z}} \varphi(G_{\theta}(y,z)) 
d(\nu \times \eta)(y,z)\right] \nonumber \\
& = \max \limits_{\psi \in Lip_{1}(\mathcal{Y})} 
\left[\int \limits_{\mathcal{Y}} \psi(y) d\nu(y) - 
\int \limits_{\mathcal{X}} \psi(F_{\phi}(x)) d\mu(x)\right] .
\end{align}


Substituting (\ref{relation_xyz_to_yx_domain_final}) in 
(\ref{OT_GAN_loss}), following relation can be 
deduced: 
\begin{align}
\label{final_GAN_loss}
&\quad \quad \quad 
l_{GAN}(G_{\theta}, F_{\phi}, E_{\omega}; \varphi, \psi, \xi)  \nonumber \\
&=\frac{1}{3} \left\{2 l_{GAN}(F_{\phi}; \psi) + l_{GAN}(E_{\omega}; \xi)\right \}
\end{align}
which proves the claim of \textit{Theorem \ref{propose}}. 

\par The above analysis 
eliminates the need of additional discriminator $\varphi$ and generator 
$G_{\theta}$ if ${G_{\theta}}_{y}$ and 
${G_{\theta}}^{z}$ are considered to be the inverse operators of forward 
operators $E_{\omega}$ and $F_{\phi}$, respectively. 
From the optimization perspective, if the normalizing constants in (\ref{final_GAN_loss}) 
are not taken into consideration, then the model requires two generators $F_{\phi}$ and 
$E_{\omega}$, and two discriminators $\psi$ and $\xi$. As the discriminators $\psi$ and 
$\xi$ discriminate between real and fake/generated $\mathcal{Y}$ domain data, and 
real and generated $\mathcal{Z}$ 
space data, respectively, instead of two discriminators, one discriminator $\psi_{\xi}$ 
is sufficient, which takes inputs in the form of 
pairs $(y, z)$. The discriminator $\psi_{\xi} (y, z)$ discriminates 
between 
the real pair $(y, z)$, where $y \sim \mathcal{Y}$ and $z \sim \mathcal{Z}$, and the 
generated/fake 
pair $(y, z)$, where $y \sim P_{F_{\phi}}(y \mid x)$ and $z \sim P_{E_{\omega}}(z \mid x)$ 
corresponding to the outputs generated through $F_{\phi}$ and $E_{\omega}$, respectively.
As a histological 
image $x$ can be generated through the invertible generator ${G_{\theta}}^{z}$ for 
a given embedding information $z \in \mathcal{Z}$, corresponding to a segmentation 
map $y \in \mathcal{Y}$ , ${G_{\theta}}^{z}$ can also be used for 
generating synthetic data during model training.

\begin{figure*}[h]
	\centerline{\includegraphics[width=0.9\textwidth]{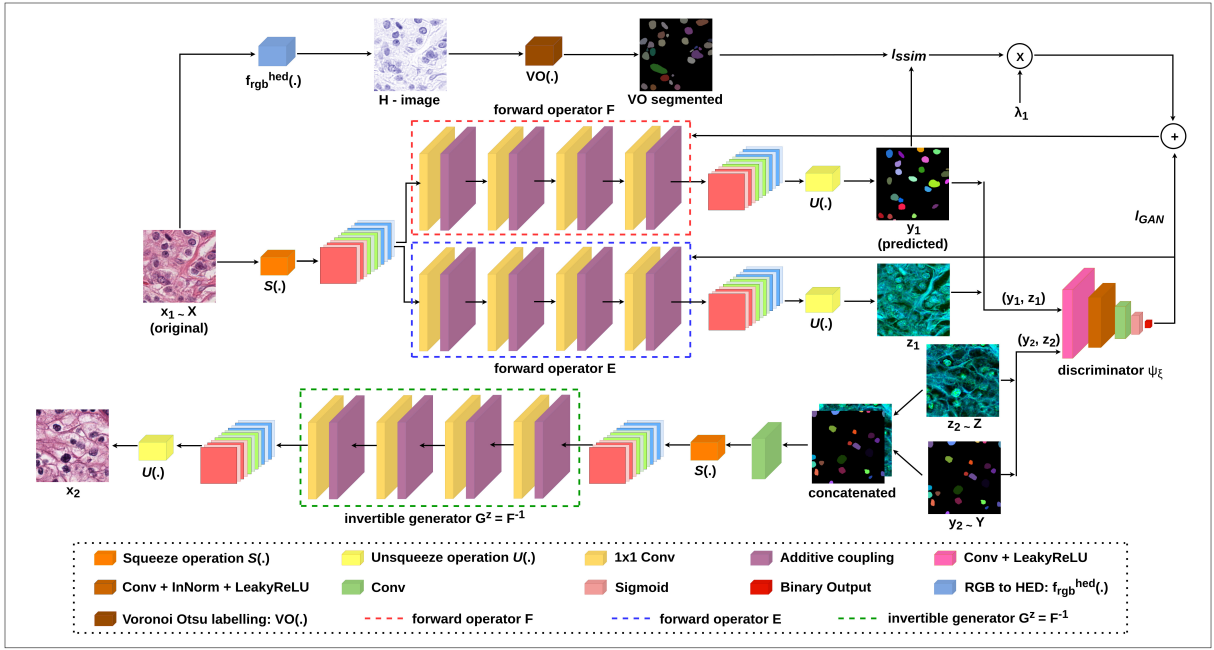}}
	\caption{Block diagram of the proposed deep generative model for 
		nuclei segmentation of histological images. Each network component is marked
		with distinct color representation. In the diagram, 'Conv' means convolutional
		layer, '$1 \times 1$ Conv' refers to invertible $1 \times 1$ convolution approach 
		proposed in \cite{Dinh2017}. Given $z_{2} \in \mathcal{Z}$, $z_{2}$ is 
		concatenated with $y_{2}$, and the concatenated image is fed to a convolution 
		layer to produce a $3$ 
		channel output. 'InNorm' represents instance normalization, 'LeakyReLU' denotes 
		leaky rectified linear unit. Here, forward operators $F$, $E$ and invertible 
		generator $G^{z}$ refer to generators $F_{\phi}$, $E_{\omega}$ and 
		${G_{\theta}}^{z}$, respectively.}
	\label{block_diagram}
	\vspace*{-0.2cm}
\end{figure*}

\subsection{Model Structure} 
Fig. \ref{block_diagram} presents the block diagram of the proposed deep generative 
model for nuclei segmentation from histological images. The forward operator $F_{\phi}$ 
generates segmentation map corresponding to each input histological image and generator 
$E_{\omega}$ embeds additional information while translating from information-rich 
image space $\mathcal{X}$ to information-poor segmentation map domain $\mathcal{Y}$. 
The discriminator ${\psi}_{\xi}$ discriminates 
between real and generated/fake pairs of segmentation map and embedding information. 
Each training image patch 
$x \in \mathcal{X}$ is fed into both forward operators $F_{\phi}(x; \Theta_{\phi})$ 
and $E_{\omega}(x; \Theta_{\omega})$, which output corresponding segmentation map $y$ 
and embedding information $z$, respectively. 
Without loss of generality, $\mathcal{Z}$ is assumed to follow 
standard normal distribution. Here, $\Theta_{\phi}$, $\Theta_{\omega}$ and 
$\Theta_{{\psi}_{\xi}}$ denote the network parameters associated with the networks 
$F_{\phi}$, 
$E_{\omega}$ and ${\psi}_{\xi}$, respectively. As ${G_{\theta}}^{z}$ is designed to be the 
inverse of the forward operator $F_{\phi}$, ${G_{\theta}}^{z}$ uses the same parameter set 
$\Theta_{\phi}$ as that of $F_{\phi}$.  
\par The overall objective of the proposed model is to generate segmentation map
corresponding to each input histological image by handling the 
information-asymmetry between the two domains. The following min-max objective 
function $J_{adv}$ is to be optimized to achieve the aforementioned objective: 
\begin{align}
\label{objective}
J_{adv} = J({\psi}_{\xi}) + J(F_{\phi}, E_{\omega})
\end{align}
where $J({\psi}_{\xi})$ is the maximization objective, attributed by the discriminator 
${\psi}_{\xi}$, and is given by
\begin{align}
J({\psi}_{\xi}) = \max \limits_{{\psi}_{\xi}} \{l_{GAN}(F_{\phi}, E_{\omega}; {\psi}_{\xi})\}
\end{align}
and $J(F_{\phi}, E_{\omega})$ is the minimization objective, attributed by the 
two generators $F_{\phi}$ and $E_{\omega}$, which is given by
\begin{align}
&\quad J(F_{\phi}, E_{\omega}) = \min \limits_{\phi, \omega} 
\{l_{GAN}(F_{\phi}, E_{\omega}; {\psi}_{\xi}) + 
\lambda_{1} l_{ssim}(F_{\phi})\} \nonumber \\
&\quad \quad \quad = \min \limits_{\phi}  \{l_{GAN}(F_{\phi}, E_{\omega}; {\psi}_{\xi}) 
+ \lambda_{1} l_{ssim}(F_{\phi})\} \nonumber \\
& \quad \quad \quad \quad \quad + \min \limits_{\omega} \{l_{GAN}(F_{\phi}, E_{\omega}; {\psi}_{\xi})\}
\end{align}
where $l_{GAN}(F_{\phi}, E_{\omega}; {\psi}_{\xi})$ can be defined as follows:
\begin{align}
&l_{GAN}(F_{\phi}, E_{\omega}; {\psi}_{\xi})
= \max \limits_{{\psi}_{\xi}} [
\int \limits_{\mathcal{Y} \times \mathcal{Z}} {\psi}_{\xi}(y, z) 
d(\nu \times \eta)(y, z)  \nonumber \\
&\quad \quad \quad \quad \quad \quad \quad 
- \int \limits_{\mathcal{X}} {\psi}_{\xi}(F_{\phi}(x), E_{\omega}(x)) d\mu(x)].
\end{align}

\begin{figure}[h]
	\centerline{\includegraphics[width=3.6in,height=1.5in]{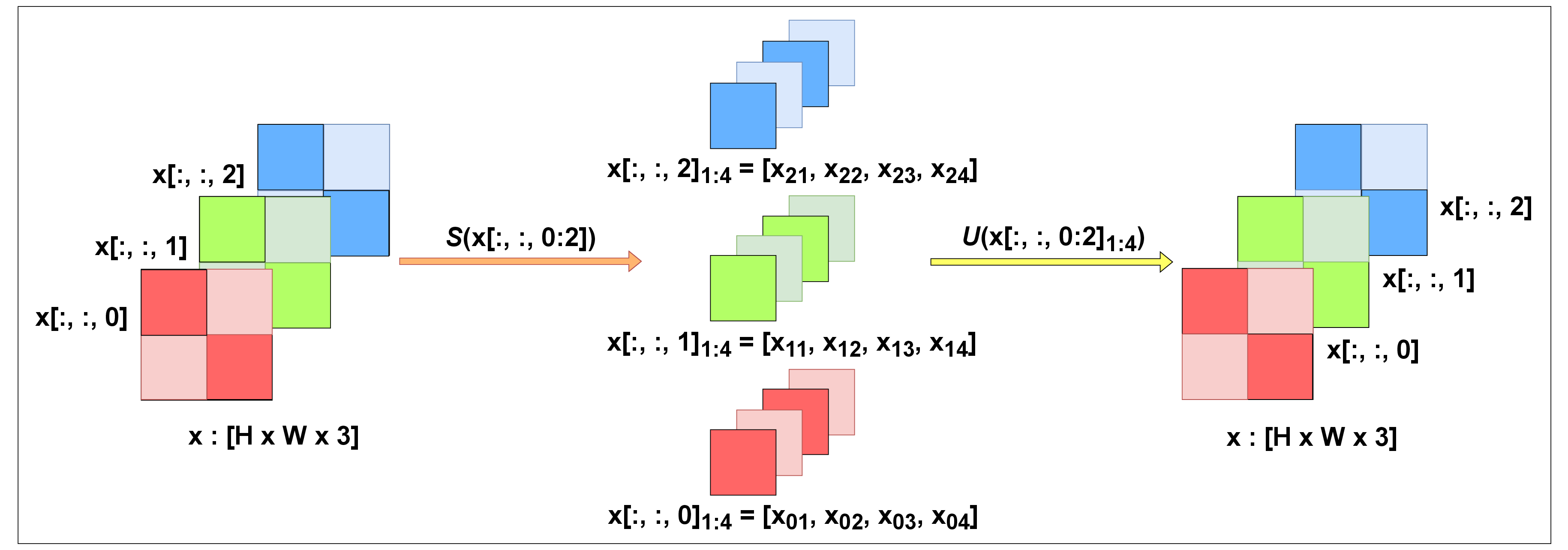}}
	\caption{Block diagram of the squeeze-unsqueeze operation. Here, $x[:, :, 0:2]$ 
		denotes $3$ channel input image. ${x[:, :, i]}_{1:4}$ represents $4$ sub-patches 
		of the $i$-th channel of input image $x \in \mathcal{X}$.}
	\label{block_squeeze_unsqueeze}
	\vspace*{-0.2cm}
\end{figure}

\par The main building block of the invertible generator is implemented using 
the concepts of invertible $1 \times 1$ convolution, proposed in \cite{Dinh2017} and 
general coupling layer, introduced in \cite{Tomczak2021}. 
Algorithm \ref{ostrich_train_algo} summarizes the training procedure 
for the proposed Ostrich model. 
To maintain the spatial 
continuity within the image patches, instead of checkerboard masking 
based squeeze operation explained in \cite{Kwon2021}, a 
simple yet effective modified squeeze-unsqueeze 
operation is introduced in this study. For an input image $x \in \mathcal{X}$, the 
squeeze operation can be mathematically represented as follows:
\begin{align*}
\left[{x[:, :, 0]}_{1:4}, {x[:, :, 1]}_{1:4}, {x[:, :, 2]}_{1:4}\right] 
= S(x[:, :, 0:2]) =  S(x)
\end{align*}
whereas the unsqueeze operation can be defined as follows:
\begin{align*}
x =  U([{x[:, :, 0:2]}_{1:4}]).
\end{align*}
The squeeze operation $S(.)$ and unsqueeze operation 
$U(.)$ are demonstrated schematically through 
Fig. \ref{block_squeeze_unsqueeze}.

\begin{algorithm*}
	\textbf{Input:} A set of training images sampled from real data distribution $P_{X}(x)$, 
	segmentation map images sampled from real data distribution $P_{Y}(y)$, 
	number of training epochs ($\tau$). \\
	\textbf{Output:} Trained parameter set $\{\Theta_{{\psi}_{\xi}}, \Theta_{F_{\phi}}, \Theta_{E_{\omega}}\}$, corresponding to three deep neural networks ${\psi}_{\xi}$, $F_{\phi}$ and $E_{\omega}$, respectively.
	\begin{algorithmic}[1]
		\For{each epoch $t = 0$ to $\tau - 1$}
		\begin{itemize}
			\item Sample mini-batch of $M$ images $\{x_{m}\}_{m=1}^{M}$ from real data
			distribution $P_{X}(x)$.
			\item Generate $M$ nuclei segmentation maps $\{y_{m}\}_{m=1}^{M}$ by using $F_{\phi}(x_{m}; \Theta_{\phi})$.
			\item Generate $M$ embedding information maps $\{z_{m}\}_{m=1}^{M}$ by using $E_{\omega}(x_{m}; \Theta_{\omega})$.
			\item Sample mini-batch of $M$ nuclei segmentation maps $\{\acute{y}_{m}\}_{m=1}^{M}$ from real segmentation map image distribution $P_{Y}(y)$.
			\item Sample mini-batch of $M$ embedding information maps $\{\acute{z}_{m}\}_{m=1}^{M}$ from embedding information prior $P_{Z}(z)$.
			\item Update discriminator ${\psi}_{\xi}$ by ascending in the direction of stochastic gradient:
			\begin{equation*}
			\nabla_{\Theta_{{\psi}_{\xi}}} \left(\frac{1}{M} \sum_{m=1}^{M}
			\left[{\psi}_{\xi}(\acute{y}_{m}, \acute{z}_{m}; \Theta_{{\psi}_{\xi}}) -
			{\psi}_{\xi}(y_{m}, z_{m}; \Theta_{{\psi}_{\xi}})
			\right]\right)
			\end{equation*}
			\item Update segmentation map generator $F_{\phi}$ by descending in the direction of stochastic gradient:
			\begin{equation*}
			\nabla_{\Theta_{F_{\phi}}}\left(\frac{1}{M} \sum_{m=1}^{M}
			\left[-{\psi}_{\xi}(y_{m}, z_{m}; \Theta_{{\psi}_{\xi}}) + \lambda_{1} 
			\lVert 1 - SSIM(y_{m}, VO(f_{rgb}^{hed}(x_{m})[:, :, 0]))\rVert \right]\right)
			\end{equation*}
			\item Update embedding map generator $E_{\omega}$ by descending in the direction of stochastic gradient:
			\begin{equation*}
			\nabla_{\Theta_{E_{\omega}}}\left(\frac{1}{M} \sum_{m=1}^{M}
			\left[-{\psi}_{\xi}(y_{m}, z_{m}; \Theta_{{\psi}_{\xi}}) \right]\right)
			\end{equation*}
		\end{itemize}
		\EndFor
		\State Stop.
	\end{algorithmic}
	\caption{Mini-batch stochastic gradient descent training of proposed Ostrich model.}
	\label{ostrich_train_algo}
\end{algorithm*}

\section{Performance Analysis}
\label{performance}
This section analyzes the performance of the proposed 
deep generative model, termed as Ostrich.

\subsection{Experimental Setup}
\label{experiments}
A brief description of the existing algorithms compared, data sets and quantitative 
indices used, and details regarding model training are provided next.

\subsubsection{Algorithms Compared}
\label{algorithms}
The performance of the proposed Ostrich model in nuclei segmentation is analyzed 
extensively and compared with that of several standard object segmentation models, 
such as U-Net (2015) \cite{Ronneberger2015}, Mask-R-CNN (2017) \cite{He2017}, U-Net++ (2018) \cite{Zhou2018}; 
and state-of-the-art nuclei segmentation methods, namely, HoVer-Net {(2019) \cite{Graham2019}, 
multi-organ nuclei segmentation (MoNS) (2020) \cite{Mahmood2020}, Stardist (2022) \cite{Weigert2022}, 
Swin-MIL (2022) \cite{Qian2022} and BoNuS (2024) \cite{Lin2024}.   

\subsubsection{Data Sets Used}
\label{datasets}
The performance of different variants of the proposed model and several state-of-the-art 
algorithms in nuclei segmentation is analyzed using the following two publicly available 
data sets.

\begin{itemize}
	\item TCGA Data \cite{Kumar2017}: This multi-organ image data set contains $30$ images of resolution $1000\times 1000$, that were cropped from nuclei-dense regions of whole slide images downloaded from The Cancer Genome Atlas (TCGA). This H\&E stained data set contains both instance segmentation and binary semantic segmentation maps corresponding to each histological image. 
	\item CoNIC Data \cite{Graham2021}: This H\&E stained image set was published as a challenge data set: Colon Nuclei Identification and Counting (CoNIC) Challenge, 2022. The data set consists of 4,981 image patches having resolution of $256\times 256$. This data set contains instance segmentation maps corresponding to each histological image.
\end{itemize} 

%

Four standard segmentation evaluation 
metrics, namely, Dice coefficient, Jaccard index, precision and recall, 
are used to analyze the performance of different algorithms.  

\subsubsection{Training Details}
\label{training}
The TCGA data set is randomly split into training, validation and test set 
with a ratio of $5:1:4$. 
Hence, $2,535$ overlapping image patches of size 
$256\times 256$ corresponding to $15$ training set images of TCGA data set 
have been used for training, whereas the validation set 
contains $507$ overlapping patches and the test set comprises of 
$192$ non-overlapping image patches corresponding to $3$ and $12$ images, respectively. 
The CoNIC data set consists 
of $4,981$ image patches, among which $2,930$ image patches have been used for training, $586$ 
patches for validation and $1,465$ images have been used for testing the performance of 
the proposed model. The incorporation of validation
set for both the data sets prevents the model from over-fitting and ensures early
stopping based on validation loss.

\subsection{Performance Analysis on TCGA Data}
\label{tcga_results}
The effectiveness of the proposed Ostrich model is validated through the 
following analyses.

\subsubsection{Ablation Study}
\label{ablationstudy}
The objective function of the proposed Ostrich model is comprised of two constituent terms: a 
GAN objective term and a regularization term. Let, the regularization term $l_{ssim}(F_{\phi})$ 
be denoted by $R$. The importance of the regularization term $R$ is analyzed by an ablation study, 
where $R$ is omitted from the objective function by setting $\lambda_{1} = 0$. The performance 
of the proposed Ostrich model in nuclei segmentation is then compared with that of the model 
in the absence of $R$, 
and the corresponding results are presented in Table \ref{index_existing_TCGA}. When $\lambda_{1}$ 
is set to $0$, the model is referred to as ``Ostrich$\setminus R$".  From the results reported in 
Table \ref{index_existing_TCGA}, it is evident that the proposed model performs better than the ``Ostrich$\setminus R$" model with respect to all 
the segmentation evaluation indices used in this study. The qualitative comparison 
presented in Fig. \ref{ablation_TCGA} ensures that, with the initial perturbation through regularization 
term $R$, the proposed model is able to capture the intrinsic details of underlying data distribution. 
The statistical significance of the proposed Ostrich model is analyzed with respect to 
p-values computed through paired-\textit{t} (one-tailed) and Wilcoxon signed-rank (one-tailed) 
tests. From the p-values reported in Table \ref{p_values_TCGA}, it can 
be observed that the proposed model performs significantly better than the ``Ostrich$\setminus R$" model, considering $95\%$ confidence level, with respect to both paired-\textit{t} and Wilcoxon signed-rank tests.

\begin{table}[h]
	\begin{center}
		\caption{Quantitative performance analysis in nuclei segmentation on TCGA Data set: State-of-the-art models vs Ostrich}
		\label{index_existing_TCGA}
		\footnotesize{
			\begin{tabular}{|c|c|c|c|c|}\hline
				Methods	&	Dice 	&	Jaccard	&	Precision	&	Recall	\\\hline	
				Ostrich	&	\textbf{0.788035}	&	\textbf{0.641520}	&	\textbf{0.813523}	&	0.764096	\\\hline \hline	
				Ostrich$\setminus R$	&	0.761814	&	0.606573	&	0.791683	&	0.734116	\\\hline \hline
				CycleGAN	&	0.754221	&	0.598782	&	0.764761	&	0.743967	\\	
				OT-CycleGAN	&	0.766668	&	0.614903	&	0.804106	&	0.732561	\\	
				Asym-CycleGAN	&	0.783812	&	0.638815	&	0.806971	&	0.761945	\\\hline \hline
				Checkerboard	&	0.772020	&	0.629522	&	0.785028	&	0.759437	\\\hline \hline
				U-Net (2015)	&	0.645650	&	0.446128	&	0.624558	&	0.668216	\\	
				Mask-R-CNN (2017)	&	0.747086	&	0.585659	&	0.799975	&	0.700757	\\	
				U-Net++ (2018) &	0.773155	&	0.620294	&	0.796708	&	0.750955	\\	
				HoVer-Net (2019)	&	0.744516	&	0.586780	&	0.811804	&	0.687529	\\	
				MoNS (2020)	&	0.755509	&	0.614706	&	0.748012	&	0.763157	\\	
				Stardist (2022)	&	0.743228	&	0.585713	&	0.811579	&	0.685495	\\		
				Swin-MIL (2022)	&	0.749044	&	0.593266	&	0.801989	&	0.702656	\\	
				BoNuS (2024)	&	0.784740	&	0.639496	&	0.806351	&	\textbf{0.764257}	\\\hline	
		\end{tabular}}
	\end{center}
\end{table}

\begin{figure}[h]
	\centerline{\includegraphics[width=2.4in,height=1.4in]{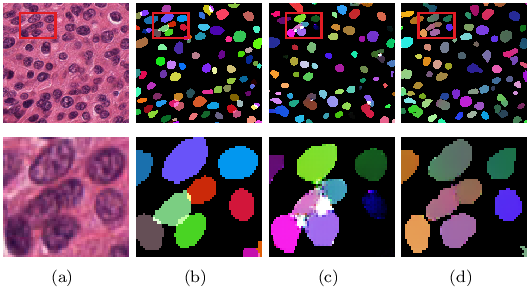}}
	\caption{Qualitative performance analysis in nuclei instance segmentation: 
		(a) Original image, (b) Ground-truth; Segmentation performance of (c) Ostrich$\setminus R$ 
		and (d) Ostrich. Top row presents marked-up images from TCGA 
		data, and bottom row presents the zoomed-in region from the marked-up patches.}
	\label{ablation_TCGA}
\end{figure}

\begin{table}[h]
	\begin{center}
		\caption{Statistical significance analysis of different algorithms with respect to 
			Paired-\textit{t} (one-tailed) and Wilcoxon signed rank (one-tailed) tests on TCGA Data set}
		\label{p_values_TCGA}
		\footnotesize{
			\begin{tabular}{|@{}c@{}|@{}c@{}|c|c|c|c|}\hline
				Methods	& Test &	Dice 	&	Jaccard	&	Precision	&	Recall	\\\hline	\hline
				Ostrich$ \setminus R$	& \multirow{14}*{{\rotatebox[origin=c]{90}{Paired-\textit{t}}}}	&	9.88E-30	&	3.41E-30	&	1.44E-09	&	1.30E-10	\\\cline{1-1} \cline{3-6}
				CycleGAN	&	&	4.88E-32	&	3.39E-34	&	3.69E-32	&	8.40E-10	\\	
				OT-CycleGAN	&	&	2.90E-33	&	2.26E-33	&	1.04E-05	&	6.73E-21	\\	
				Asym-CycleGAN	&	&	\textit{6.49E-02}	&	\textit{5.42E-02}	&	5.43E-05	&	\textit{8.31E-02}	\\\cline{1-1}	\cline{3-6}
				Checkerboard	&	&	4.66E-13	&	5.70E-14	&	1.83E-61	&	1.10E-02	\\\cline{1-1}	\cline{3-6}
				U-Net (2015)	&	&	6.67E-30	&	4.18E-36	&	1.25E-76	&	6.68E-06	\\	
				Mask-R-CNN (2017)	&	&	1.17E-37	&	4.55E-39	&	3.77E-05	&	1.28E-25	\\	
				U-Net++ (2018)	&	&	1.08E-17	&	3.86E-18	&	5.03E-09	&	7.46E-03	\\	
				HoVer-Net (2019)	&	&	1.77E-41	&	7.41E-45	&	\textit{2.37E-01}	&	7.65E-54	\\	
				MoNS (2020)	&	&	1.10E-14	&	1.84E-15	&	3.34E-48	&	\textit{1.21E-01}	\\	
				Stardist (2022)	&	&	1.46E-24	&	8.26E-26	&	\textit{3.09E-01}	&	9.93E-27	\\		
				Swin-MIL (2022)	&	&	1.95E-37	&	1.64E-41	&	1.74E-04	&	3.25E-49	\\	
				BoNuS (2022)	&	&	\textit{1.42E-01}	&	\textit{1.05E-01}	&	9.20E-07	&	\textit{9.43E-01}	\\\hline	\hline
				Ostrich $ \setminus R$	& \multirow{14}*{{\rotatebox[origin=c]{90}{Wilcoxon}}}	&	2.37E-23	&	2.34E-23	&	5.57E-07	&	2.01E-09	\\\cline{1-1}	\cline{3-6}
				CycleGAN	&	&	9.05E-27	&	6.08E-27	&	1.66E-30	&	7.70E-10	\\	
				OT-CycleGAN	&	&	8.58E-25	&	6.83E-25	&	7.49E-04	&	6.37E-18	\\	
				Asym-CycleGAN	&	&	\textit{8.56E-02}	&	\textit{7.80E-02}	&	1.13E-03	&	\textit{7.67E-02}	\\\cline{1-1}	\cline{3-6}
				Checkerboard	&	&	7.74E-13	&	2.49E-13	&	8.57E-33	&	1.21E-02	\\\cline{1-1}	\cline{3-6}
				U-Net (2015)	&	&	8.61E-32	&	7.06E-32	&	1.89E-33	&	4.45E-02	\\	
				Mask-R-CNN (2017)	&	&	2.74E-28	&	3.20E-28	&	1.55E-03	&	1.41E-20	\\	
				U-Net++ (2018)	&	&	9.70E-17	&	6.71E-17	&	2.86E-07	&	1.88E-03	\\	
				HoVer-Net (2019)	&	&	1.63E-30	&	1.23E-30	&	\textit{6.85E-02}	&	4.13E-32	\\	
				MoNS (2020)	&	&	1.46E-13	&	3.65E-14	&	2.11E-33	&	\textit{1.10E-01}	\\	
				Stardist (2022)	&	&	7.73E-22	&	6.91E-22	&	\textit{3.40E-01}	&	3.03E-23	\\		
				Swin-MIL (2022)	&	&	9.15E-32	&	8.48E-32	&	3.83E-02	&	1.47E-32	\\	
				BoNuS (2024)	&	&	\textit{8.18E-02}	&	\textit{5.86E-02}	&	7.36E-06	&	\textit{9.00E-01}	\\\hline	
		\end{tabular}}
	\end{center}
\end{table}

\subsubsection{Comparison with Baseline Models}
\label{baseline}

\begin{table}[h]
	\begin{center}
		\caption{Number of parameters involved in each of the baseline I2I translation models and Ostrich (in millions)}
		\label{param_baseline}
		\footnotesize{
			\begin{tabular}{|c|c|}\hline
				Methods	&	No. of Parameters (M)	\\\hline
				CycleGAN	&	18.035	\\
				OT-CycleGAN	&	4.353	\\
				Asym-CycleGAN	&	27.053	\\
				Ostrich	&	7.271	\\\hline
		\end{tabular}}
	\end{center}
\end{table}

The performance of the proposed Ostrich model is compared with that of two existing 
I2I translation models, namely, CycleGAN \cite{Zhu2017} and cycle-free CycleGAN \cite{Kwon2021}. 
To establish the significance of the proposed optimal transport and measure theory based 
analysis, the existing CycleGAN model is also extended by considering an embedding space for capturing 
additional information while translating from information-rich histological image space to 
information-poor segmentation map domain. This asymmetric CycleGAN model is comprised of three generators 
and three discriminators and is referred to as Asym-CycleGAN in this section. For simplicity, the 
optimal transport driven cycle-free CycleGAN model is referred to as OT-CycleGAN. 
Similar to the proposed Ostrich model, the regularization term $l_{ssim}(F_{\phi})$ is also 
incorporated in all three models, CycleGAN, OT-CycleGAN and Asym-CycleGAN for this comparative analysis.
From the results reported in Table \ref{index_existing_TCGA}, it can be observed that the proposed 
Ostrich model outperforms baseline I2I translation models, with respect to all the segmentation evaluation 
indices, namely, Dice, Jaccard, precision and recall. 
From the p-values reported in Table \ref{p_values_TCGA}, it can be noted 
that the Ostrich model performs significantly better than CycleGAN and OT-CycleGAN, and 
better than Asym-CycleGAN model, but not significantly. 
It is also evident from Table \ref{param_baseline} that the proposed Ostrich model uses less 
number of parameters 
compared to CycleGAN and Asym-CycleGAN models. Though OT-CycleGAN uses less number 
of parameters 
than the proposed Ostrich model, it does not capture additional information to balance information 
asymmetry between two image domains, and eventually fails to outperform the proposed model. 
Through the qualitative representation 
presented in Fig. \ref{baseline_TCGA}, it can also be observed that the proposed model performs 
better than all the aforementioned baseline models.  

\begin{figure}[h]
	\centerline{\includegraphics[width=3.45in,height=1.3in]{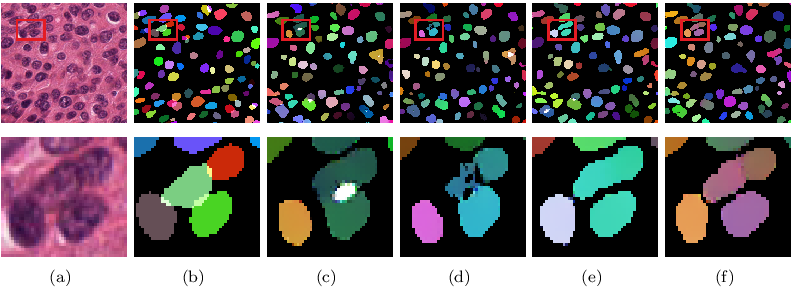}}
	\caption{Performance analysis in nuclei segmentation: 
		(a) Original histological image, (b) Ground-truth segmentation map; segmentation performance of (c) CycleGAN , (d) OT-CycleGAN, (e) Asym-CycleGAN and (f) Ostrich. 
		Top row presents marked-up images from TCGA data, and bottom row presents the zoomed-in region from the marked-up image patches.}
	\label{baseline_TCGA}
\end{figure}     

\subsubsection{Importance of Spatially-Constrained Squeeze Operation}
\label{squeeze-unsqueeze}
The performance of the proposed Ostrich model with spatially-constrained 
squeeze operation is compared with that of the proposed model with checkerboard 
masking based squeeze operation \cite{Kwon2021}, and the corresponding 
results are reported in Table \ref{index_existing_TCGA}. The values reported in 
Table \ref{index_existing_TCGA} ensure the fact that spatially-constrained 
squeeze operation based proposed Ostrich model 
outperforms the model with checkerboard masking based squeeze operation, with 
respect to all the segmentation evaluation indices. 
The qualitative comparison of the 
performance of the proposed model and checkerboard masking based model is 
presented in Fig. \ref{squeezeblock_TCGA}. It can be observed in 
Fig. \ref{squeezeblock_TCGA} that the proposed Ostrich model performs better 
than the checkerboard masking based model in separating overlapping nuclei 
structures. This is due to the fact that the proposed model can capture spatial 
connectivity through spatially-constrained squeeze operation, which checkerboard 
masking based model fails to capture. The p-values reported in 
Table \ref{p_values_TCGA} establish the fact that the proposed Ostrich model with 
spatially-constrained squeeze operation performs significantly better than the 
checkerboard masking based counterpart. 

\begin{figure}[h]
	\centerline{\includegraphics[width=2.5in,height=1.35in]{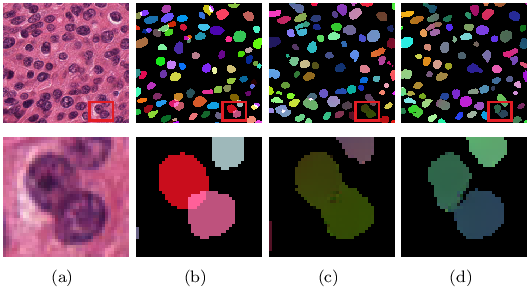}}
	\caption{Performance analysis of nuclei segmentation: 
		(a) Original histological image, (b) Ground-truth; segmentation performance of models with
		(c) checkerboard masking based and (d) spatially-constrained 
		squeeze operation. Top row presents marked-up images from TCGA data, and bottom row presents the zoomed-in region from the marked-up image patches.}
	\label{squeezeblock_TCGA}
\end{figure} 

\subsubsection{Performance in Binary Semantic Segmentation}
\label{performance_semantic}

The performance of the proposed Ostrich model on TCGA data set in binary semantic 
segmentation is validated through the qualitative 
results presented in Fig. \ref{binary_TCGA}. It presents two image patches, 
their corresponding ground-truth images and two binary segmentation maps predicted 
by the proposed Ostrich model. From the qualitative analysis, it is clear that the 
proposed model can capture heterogeneous nuclei shapes. It is possible due to the 
fact that the proposed model does not assume the shapes of the nuclei objects 
apriori. The results reported in Table \ref{index_semantic} also ensure the fact 
that the proposed Ostrich model performs well in binary semantic segmentation.

\begin{figure}[h]
\centerline{\includegraphics[width=3.45in,height=1.3in]{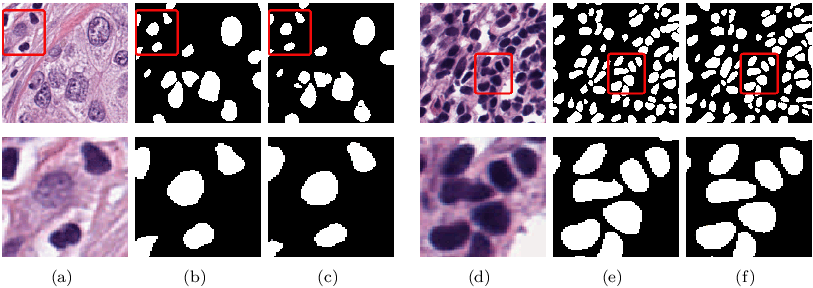}}
\caption{Performance analysis in binary semantic segmentation: (a) histological 
image patch 1, (b) ground-truth segmentation map 1, (c) predicted map 1; 
(d) image patch 2, (e) ground-truth segmentation 2 and (f) predicted map 2. 
Top row presents marked-up images from TCGA data, and 
bottom row presents the zoomed-in region from the marked-up image patches.}
\label{binary_TCGA}
\end{figure}

\begin{table}[h]
	\begin{center}
		\caption{Performance in binary semantic segmentation}
		\label{index_semantic}
		\footnotesize{
			\begin{tabular}{|c|c|c|c|c|}\hline
				Model	&	Dice 	&	Jaccard	&	Precision	&	Recall	\\\hline
				Ostrich	&	0.810099	&	0.746830	&	0.817761	&	0.802580	\\\hline
		\end{tabular}}
	\end{center}
\end{table}

\subsubsection{Comparison with State-of-the-Art Methods}

The performance of the proposed Ostrich model is finally compared with that of several 
state-of-the-art nuclei segmentation approaches on TCGA data 
set and the results with respect to different quantitative indices are reported in Table 
\ref{index_existing_TCGA}. From the results 
reported in Table \ref{index_existing_TCGA}, it can be observed that, in case of TCGA data, 
the proposed model outperforms all the state-of-the-art nuclei segmentation approaches, with respect 
to three segmentation evaluation indices, namely, Dice, Jaccard and precision. With respect to 
recall score, BoNuS performs slightly better than the proposed Ostrich model. 
The qualitative performance of the Ostrich model, along with a comparison with 
state-of-the-art approaches, in nuclei segmentation, is presented in Fig. \ref{existing_TCGA} 
using two image patches from the TCGA data set. 
Analyzing the p-values reported in
Table \ref{p_values_TCGA}, it is evident that, with respect both paired-\textit{t} and 
Wilcoxon signed-rank tests, 
the proposed Ostrich model performs significantly better in $29$ out of $36$ cases, better 
but not significantly in $6$ cases. 

\begin{figure*}[h]
	\centerline{\includegraphics[width=0.99\textwidth]{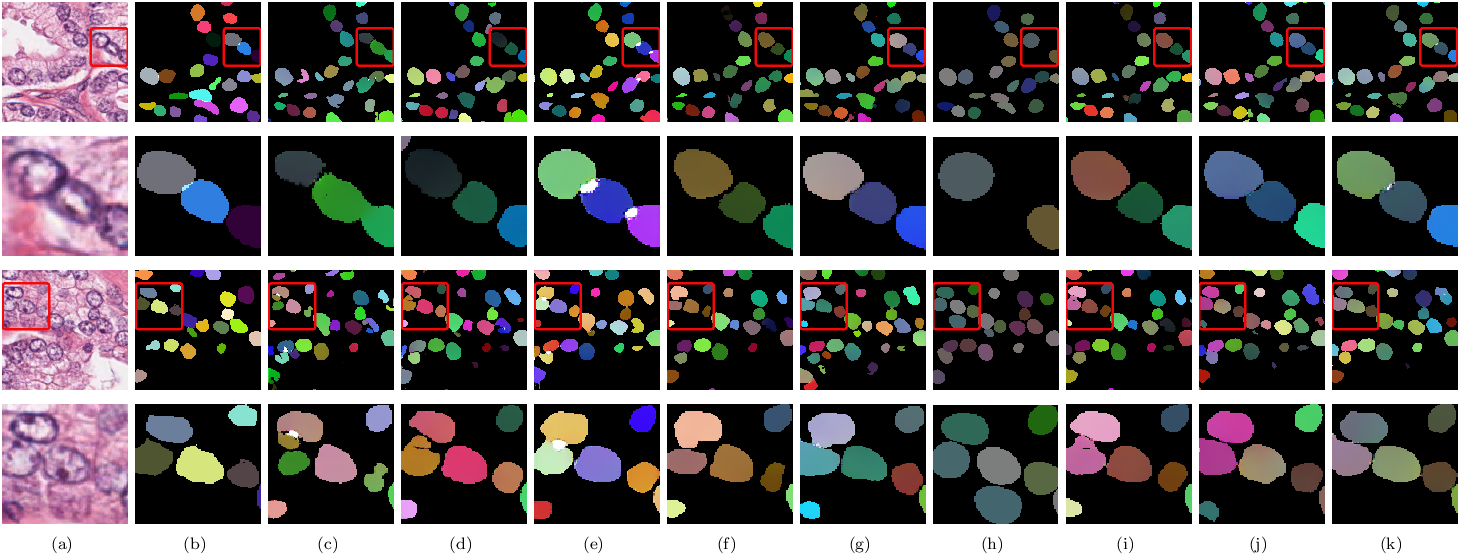}}
	\caption{(a) Original images of TCGA data, (b) ground-truth segmentation map; and segmentation maps obtained using existing nuclei segmentation methods:
		(c) U-Net (2015), (d) Mask-R-CNN (2017), (e) U-Net++ (2018), (f) HoVer-Net (2019), (g) MoNS (2020), (h) Stardist (2022), (i) Swin-MIL (2022), (j) BoNuS (2024) and (k) Ostrich. 
		Row 1 and row 3 present two marked-up images from TCGA data, 
		and row 2 and row 4 present the zoomed-in regions corresponding to the marked-up 
		image patches in row 1 and row 3, respectively.}
	\label{existing_TCGA}
\end{figure*}

\subsection{Performance Analysis on CoNIC Data}

Finally, the performance of the proposed Ostrich model is compared with that of existing 
nuclei segmentation methods on the CoNIC data set and the corresponding results are reported 
in Table \ref{index_existing_conic}. 
From the results provided in Table \ref{index_existing_conic}, it can be easily observed 
that, in case of CoNIC data set, the proposed Ostrich model outperforms existing nuclei 
segmentation approaches with respect all the evaluation indices. 
The high precision and low recall values of Stardist ensure the fact that it is confident about 
a few positive instances it predicts, but misses a substantial number of true positive instances. 
The segmentation performance 
of the proposed Ostrich model is also validated through a qualitative comparison presented 
in Fig. \ref{existing_conic}. The qualitative comparison through zoomed-in regions of the 
segmentation maps makes it clear that the proposed Ostrich model captures the nuclei structures 
better than most of the existing approaches.
Finally, analyzing the p-values reported in
Table \ref{p_values_conic}, it can be noted that 
the proposed Ostrich model performs significantly better in $34$ out of $36$ cases, better 
but not significantly in $2$ cases, with respect paired-\textit{t} test. From the p-values 
reported in Table \ref{p_values_conic}, 
it can also be observed that, irrespective of paired-\textit{t} and 
Wilcoxon signed-rank tests, the proposed Ostrich model performs significantly better in $34$ out 
of $36$ cases, better but not significantly in just $2$ cases.
In Table \ref{p_values_conic}, the p-values less 
than or equal to 9.99E-99 are reported as 9.99E-99. 

\begin{table}[h]
	\begin{center}
		\caption{Quantitative performance analysis in nuclei segmentation on CoNIC Data set: State-of-the-art models vs Ostrich}
		\label{index_existing_conic}
		\footnotesize{
			\begin{tabular}{|c|c|c|c|c|}\hline
				Methods	&	Dice	&	Jaccard	&	Precision	&	Recall	\\\hline	
				Ostrich	&	\textbf{0.741107}	&	\textbf{0.591310}	&	\textbf{0.769381}	&	\textbf{0.793093}	\\\hline	\hline
				U-Net (2015)	&	0.715636	&	0.561178	&	0.704462	&	0.681140	\\	
				Mask-R-CNN (2017)	&	0.717821	&	0.562671	&	0.725431	&	0.706849	\\	
				U-Net++ (2018) &	0.721316	&	0.568022	&	0.721591	&	0.706789	\\	
				HoVer-Net (2019)	&	0.740318	&	0.590560	&	0.747550	&	0.698034	\\	
				MoNS (2020)	&	0.722989	&	0.568692	&	0.722632	&	0.733005	\\	
				Stardist (2022)	&	0.374534	&	0.240549	&	0.716413	&	0.275528	\\		
				Swin-MIL (2022)	&	0.715991	&	0.560825	&	0.764054	&	0.682995	\\	
				BoNuS (2024)	&	0.724678	&	0.571128	&	0.753906	&	0.768233	\\\hline	
		\end{tabular}}
	\end{center}
\end{table}

\begin{figure*}[h]
	\centerline{\includegraphics[width=0.99\textwidth]{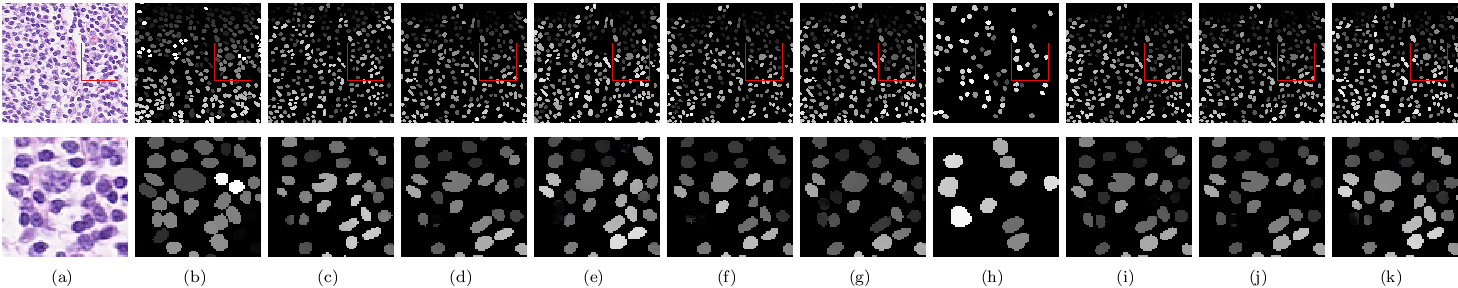}}
	\caption{(a) Original images of CoNIC data, (b) ground-truth segmentation map; and segmentation maps 
		obtained using different nuclei segmentation methods:
		(c) U-Net (2015), (d) Mask-R-CNN (2017), (e) U-Net++ (2018), (f) HoVer-Net (2019), (g) MoNS (2020), (h) Stardist (2022), (i) Swin-MIL (2022), (j) BoNuS (2024) and (k) Ostrich. 
		Top row presents a marked-up images from CoNIC 
		data, and bottom row presents the zoomed-in region from the marked-up patches.}
	\label{existing_conic}
\end{figure*}

\begin{table}[h]
	\begin{center}
		\caption{Statistical significance analysis of different algorithms with respect to 
			Paired-\textit{t} (one-tailed) and Wilcoxon signed rank (one-tailed) tests on CoNIC data}
		\label{p_values_conic}
		\footnotesize{
			\begin{tabular}{|@{}c@{}|@{}c@{}|c|c|c|c|}\hline
				Methods	& Test &	Dice	&	Jaccard	&	Precision	&	Recall	\\\hline	
				U-Net (2015)	& \multirow{9}*{{\rotatebox[origin=c]{90}{Paired-\textit{t}}}}	&	1.21E-85	&	1.27E-86	&	9.99E-99	&	9.99E-99  \\	
				Mask-R-CNN (2017)	&	&	9.99E-99	&	1.11E-99	&	9.99E-99	&	9.99E-99	\\	
				U-Net++ (2018)	&	&	1.02E-63	&	2.46E-87	&	9.99E-99	&	9.99E-99	\\	
				HoVer-Net (2019)	&	&	\textit{8.81E-02}	&	\textit{1.44E-01}	&	4.57E-98	&	9.99E-99	\\	
				MoNS (2020)	&	&	9.99E-99	&	9.99E-99	&	9.99E-99	&	9.99E-99	\\	
				Stardist (2022)	&	&	9.99E-99	&	1.11E-99	&	1.24E-82	&	9.99E-99	\\		
				Swin-MIL (2022)	&	&	9.99E-99	&	9.99E-99	&	5.75E-20	&	9.99E-99	\\	
				BoNuS (2024)	&	&	4.28E-78	&	1.12E-93	&	9.99E-99	&	9.99E-99	\\\hline	\hline
				U-Net (2015)	& \multirow{9}*{{\rotatebox[origin=c]{90}{Wilcoxon}}}	&	1.17E-71	&	2.67E-71	&	9.99E-99	&	9.99E-99	\\	
				Mask-R-CNN (2017)	&	&	9.99E-99	&	9.99E-99	&	9.99E-99	&	9.99E-99	\\	
				U-Net++ (2018)	&	&	9.99E-99	&	9.99E-99	&	9.99E-99	&	9.99E-99	\\	
				HoVer-Net (2019)	&	&	\textit{1.93E-01}	&	\textit{2.56E-01}	&	9.99E-99	&	9.99E-99	\\	
				MoNS (2020)	&	&	9.99E-99	&	9.99E-99	&	9.99E-99	&	9.99E-99	\\	
				Stardist (2022)	&	&	9.99E-99	&	1.11E-99	&	1.08E-83	&	9.99E-99	\\		
				Swin-MIL (2022)	&	&	9.99E-99	&	9.99E-99	&	2.37E-31	&	9.99E-99	\\	
				BoNuS (2024)	&	&	9.99E-99	&	9.99E-99	&	9.99E-99	&	9.99E-99	\\\hline	
		\end{tabular}}
	\end{center}
\end{table}

\par The quantitative results reported in Table \ref{index_existing_TCGA} and 
Table \ref{index_existing_conic}, and the qualitative analyses presented in Fig. \ref{existing_TCGA} 
and Fig. \ref{existing_conic} ensure the fact that the proposed Ostrich model outperforms 
the state-of-the-art nuclei segmentation approaches in almost all the cases.  
The proposed model performs consistently 
well in different types of nuclei segmentation tasks due to the following facts: (a) it employs 
I2I translation 
paradigm, which is invariant to different target domain representations, and (b) incorporation 
of an additional embedding space helps the proposed model to balance information disparity 
between two asymmetric image domains.      

\section{Conclusion}
\label{conclusion}
Nuclei segmentation is one of the fundamental and critical steps 
in histological image analysis as it aids in subsequent analysis of 
tissue specimens to estimate tumor cellularity, which is very 
crucial in the diagnosis and prognosis of cancer patients. In this respect, 
the main contributions of the current study are listed as follows:

\begin{itemize}
	\item development of a deep generative model for segmenting nuclei 
	regions from histological images, which considers an embedding space 
	to handle information-imbalance between two image domains; 
	\item  development of an invertible generator, which achieves 
	asymmetric I2I translation with lower network complexity;
	\item introduction of a regularization term to perturb image 
	generation towards the accurate direction at the initial stage of training; and  
	\item demonstrating the effectiveness of the proposed model, along
	with a comparison with existing algorithms, on
	publicly available H\&E stained histology image data sets.
\end{itemize}

The proposed model utilizes I2I translation paradigm, which 
enhances the applicability of the model in different 
types of nuclei segmentation tasks. From the quantitative and qualitative 
analysis presented in this paper, it is evident that the proposed model performs 
consistently well in both binary semantic segmentation and instance segmentation 
tasks with different types of target domain image representations. The proposed 
model does not consider any prior information about the shape of nuclei 
structures, which helps the model to capture heterogeneous nuclei shapes 
from tissue images belonging to different organs. 

\section*{Appendix}
This section presents the derivation of the dual formulation corresponding 
to the primal form presented in (\ref{OT_Primal_Cost}):
\begin{align}
\label{appendix:OT_Primal_Cost}
\inf_{\pi \in \Pi(\mu, \nu, \eta)} 
\int \limits_{\mathcal{X} \times \mathcal{Y} \times \mathcal{Z}} 
\mathpzc{C}(x, y, z; G_{\theta}, F_{\phi}, E_{\omega}) d\pi(x, y, z)
\end{align} 
where the transportation cost 
$\mathpzc{C}(x, y, z; G_{\theta}, F_{\phi}, E_{\omega})$ is defined in 
(\ref{OT_Primal_with_SSIM}).
So, (\ref{appendix:OT_Primal_Cost}) can be reformulated as follows:
\begin{align}
\label{appendix:OT_Primal_parts}
\inf_{\pi \in \Pi(\mu, \nu, \eta)} 
\int \limits_{\mathcal{X} \times \mathcal{Y} \times \mathcal{Z}} 
\mathpzc{\hat{C}}(x, y, z) d\pi(x, y, z) + 
\lambda_{1} l_{ssim}(F_{\phi})
\end{align}
where $l_{ssim}(F_{\phi})$ is defined in (\ref{regularization_term}) and 
\begin{align}
\label{appendix:OT_Primal_main}
&\mathpzc{\hat{C}}(x, y, z) = \lVert x - G_{\theta}(y, z) \rVert + 
\lVert F_{\phi}(x) - y \rVert + \lVert E_{\omega}(x) - z \rVert .
\end{align}
To compute the dual formulation of the above primal, let
\begin{align}
\label{appendix:OT_Primal_main_define}
K(G_{\theta}, F_{\phi}, E_{\omega}) = 
\min_{\pi \in \Pi(\mu, \nu, \eta)} \int \limits_{\mathcal{X} \times 
	\mathcal{Y} \times \mathcal{Z}} 
\mathpzc{\hat{C}}(x, y, z)  d\pi(x, y, z) .
\end{align}
Let $\pi^{*}$ be the optimal joint measure, that is, 
$\pi^{*} = \arg \min \limits_{\pi} \{\pi \in \Pi(\mu, \nu, \eta)\}$, then 
(\ref{appendix:OT_Primal_main_define}) can be rewritten as follows:
\begin{equation}
\label{appendix:OT_Primal_main_minimizer}
K(G_{\theta}, F_{\phi}, E_{\omega}) = 
\int \limits_{\mathcal{X} \times \mathcal{Y} \times \mathcal{Z}} 
\mathpzc{\hat{C}}(x, y, z)  d\pi^{*}(x, y, z) .
\end{equation}
Using (\ref{appendix:OT_Primal_main}), Kantorovich duality theorem 
\cite{Villani2008} and c-transforms of Kantorovich potentials $\varphi$, $\psi$, 
and $\xi$, (\ref{appendix:OT_Primal_main_minimizer}) can be rewritten 
as follows:
\begin{align}
\label{appendix:OT_Primal_to_Dual}
&K(G_{\theta}, F_{\phi}, E_{\omega}) = 
\int \limits_{\mathcal{X} \times \mathcal{Y} \times \mathcal{Z}} 
\{\lVert x - G_{\theta}(y, z) \rVert + 
\lVert F_{\phi}(x) - y \rVert \nonumber \\ 
&\quad \quad \quad \quad \quad \quad \quad \quad \quad \quad \quad
+ \lVert E_{\omega}(x) - z \rVert\}d\pi^{*}(x, y, z) \nonumber \\
&\quad \quad = \frac{1}{3} \{
\max \limits_{\varphi} 
[\int \limits_{\mathcal{X}} \varphi(x) d\mu(x) +
\int \limits_{\mathcal{Y} \times \mathcal{Z}} \inf \limits_{x} 
[\lVert x - G_{\theta}(y, z) \rVert \nonumber \\
&\quad \quad + \lVert F_{\phi}(x) - y \rVert +   
\lVert E_{\omega}(x) - z \rVert - \varphi(x)] 
d(\nu \times \eta)(y,z)] \nonumber \\
& + \max \limits_{\psi}
[\int \limits_{\mathcal{Y}} \psi(y) d\nu(y) + 
\int \limits_{\mathcal{X}} \inf \limits_{y} 
[\lVert x - G_{\theta}(y, z) \rVert + 
\lVert F_{\phi}(x) - y \rVert  \nonumber \\ 
& \quad \quad \quad \quad \quad + 
\lVert E_{\omega}(x) - z \rVert - \psi(y)] 
d\mu(x)] \nonumber \\
&+ \max \limits_{\xi}
[\int \limits_{\mathcal{Z}} \xi(z) d\eta(z) + 
\int \limits_{\mathcal{X}} \inf \limits_{z} 
[\lVert x - G_{\theta}(y, z) \rVert + 
\lVert F_{\phi}(x) - y \rVert  \nonumber \\ 
&\quad \quad \quad \quad \quad 
+ \lVert E_{\omega}(x) - z \rVert - \xi(z)] d\mu(x)]
\} .
\end{align}
It can be noted from the optimization perspective that 
$\lVert E_{\omega}(x) - z \rVert$ 
and $\lVert F_{\phi}(x) - y \rVert$ are not dependant on 
$\mathcal{Y}$ and $\mathcal{Z}$ space, 
respectively. 
Now, instead of finding $\inf \limits_{x}$,  
$x = G_{\theta}(y,z) = {G_{\theta}}_{y}(z) = {G_{\theta}}^{z}(y)$ is chosen. 
Again, instead of finding $\inf \limits_{y}$ and $\inf \limits_{z}$, if 
$y = F_{\phi}(x)$, and $z = E_{\omega}(x)$, respectively, are chosen, then an 
upper bound of 
$K(G_{\theta}, F_{\phi}, E_{\omega})$ can be obtained from 
(\ref{appendix:OT_Primal_to_Dual}) as follows:
\begin{align}
\label{appendix:OT_Dual_c_transform_derive}
&\quad \quad \quad \quad \quad \quad \quad \quad \quad \quad 
K(G_{\theta}, F_{\phi}, E_{\omega}) \nonumber \\
&\le \frac{1}{3} \{ \max \limits_{\varphi} 
[\int \limits_{\mathcal{X}} \varphi(x) d\mu(x) - 
\int \limits_{\mathcal{Y} \times \mathcal{Z}} \varphi(G_{\theta}(y,z)) 
d(\nu \times \eta)(y,z)] \nonumber \\
&\quad \quad \quad + \max \limits_{\psi} 
[\int \limits_{\mathcal{Y}} \psi(y) d\nu(y) - 
\int \limits_{\mathcal{X}} \psi(F_{\phi}(x)) d\mu(x)] \nonumber \\
&\quad \quad \quad + \max \limits_{\xi} 
[\int \limits_{\mathcal{Z}} \xi(z) d\eta(z) - 
\int \limits_{\mathcal{X}} \xi(E_{\omega}(x)) d\mu(x)] \nonumber \\
&+ \int \limits_{\mathcal{Y} \times \mathcal{Z}} 
[\lVert y - F_{\phi}(G_{\theta}(y,z))\rVert + 
\lVert z - E_{\omega}(G_{\theta}(y,z))\rVert ] 
\nonumber \\
&\quad \quad \quad \quad \quad \quad \quad \quad \quad \quad \quad 
d(\nu \times \eta) (y,z) \nonumber \\
& + \int \limits_{\mathcal{X}} [\lVert x - G_{\theta}(F_{\phi}(x), z) \rVert +
\lVert x - G_{\theta}(y, E_{\omega}(x)) \rVert] d\mu(x)
\} \nonumber \\
&\quad \quad \Rightarrow K(G_{\theta}, F_{\phi}, E_{\omega}) \le 
\frac{1}{3} \{
l_{GAN}(G_{\theta}, F_{\phi}, E_{\omega}; \varphi, \psi, \xi)  \nonumber \\
& \quad \quad \quad \quad \quad \quad \quad \quad 
+ l_{cycle}(G_{\theta}, F_{\phi}, E_{\omega}) \}
\end{align} 
where $l_{GAN}(G_{\theta}, F_{\phi}, E_{\omega}; \varphi, \psi, \xi)$ and 
$l_{cycle}(G_{\theta}, F_{\phi}, E_{\omega})$ are defined in 
(\ref{OT_GAN_loss}) and (\ref{loss_cycle-consistency}), respectively. Again, 
using $1$-Lipschitz continuity of 
Kantorovich potentials $\varphi$, $\psi$, and $\xi$, the following relation 
can be obtained:
\begin{align}
\label{appendix:kantorovich_potential_varphi}
&\quad \varphi(x) - \varphi(G_{\theta}(y, z)) 
\le \lVert x - G_{\theta}(y, z) \rVert 
\le \lVert x - G_{\theta}(y, z) \rVert \nonumber \\ 
&\quad \quad \quad \quad \quad 
+ \lVert F_{\phi}(x) - y \rVert + \lVert E_{\omega}(x) - z \rVert \nonumber \\
&\quad \quad \Rightarrow - \varphi(G_{\theta}(y, z)) 
\le \lVert x - G_{\theta}(y, z) \rVert - \varphi(x) \nonumber \\
&\le \lVert x - G_{\theta}(y, z) \rVert + 
\lVert F_{\phi}(x) - y \rVert +  
\lVert E_{\omega}(x) - z \rVert - \varphi(x) .
\end{align} 
Similarly, the following relations can be obtained:
\begin{align}
\label{appendix:kantorovich_potential_psi}
&\quad \quad \quad \quad - \psi(F_{\phi}(x)) 
\le \lVert F_{\phi}(x) - y \rVert - \psi(y) \nonumber \\
&\le \lVert x - G_{\theta}(y, z) \rVert + 
\lVert F_{\phi}(x) - y \rVert +  
\lVert E_{\omega}(x) - z \rVert - \psi(y)
\end{align} 
\begin{align}
\label{appendix:kantorovich_potential_xi}
&\text{and} \quad \quad \quad  - \xi(E_{\omega}(x)) 
\le \lVert E_{\omega}(x) - z \rVert - \xi(z) \nonumber \\
&\le \lVert x - G_{\theta}(y, z) \rVert + 
\lVert F_{\phi}(x) - y \rVert +  
\lVert E_{\omega}(x) - z \rVert - \xi(z)
\end{align}
The inequalities (\ref{appendix:kantorovich_potential_varphi})-(\ref{appendix:kantorovich_potential_xi}) 
lead to the following lower bound:
\begin{align}
\label{appendix:Dual_lower_bound}
&\quad \quad K(G_{\theta}, F_{\phi}, E_{\omega}) \ge 
\frac{1}{3} l_{GAN}(G_{\theta}, F_{\phi}, E_{\omega}; \varphi, \psi, \xi) .
\end{align}
It is evident from (\ref{appendix:OT_Dual_c_transform_derive}) and 
(\ref{appendix:Dual_lower_bound}) that if 
$l_{cycle}(G_{\theta}, F_{\phi}, E_{\omega})$ vanishes, then the following 
relation can be obtained:
\begin{align} 
\label{appendix:final}
K(G_{\theta}, F_{\phi}, E_{\omega}) = 
\frac{1}{3} l_{GAN}(G_{\theta}, F_{\phi}, E_{\omega}; \varphi, \psi, \xi) .
\end{align}

From the optimization perspective, the normalizing constant $\frac{1}{3}$ 
in (\ref{appendix:final}) can be ignored.

\bibliography{Ostrich}

\begin{thebibliography}{10}
\providecommand{\url}[1]{#1}
\csname url@rmstyle\endcsname
\providecommand{\newblock}{\relax}
\providecommand{\bibinfo}[2]{#2}
\providecommand\BIBentrySTDinterwordspacing{\spaceskip=0pt\relax}
\providecommand\BIBentryALTinterwordstretchfactor{4}
\providecommand\BIBentryALTinterwordspacing{\spaceskip=\fontdimen2\font plus
\BIBentryALTinterwordstretchfactor\fontdimen3\font minus
  \fontdimen4\font\relax}
\providecommand\BIBforeignlanguage[2]{{%
\expandafter\ifx\csname l@#1\endcsname\relax
\typeout{** WARNING: IEEEtran.bst: No hyphenation pattern has been}%
\typeout{** loaded for the language `#1'. Using the pattern for}%
\typeout{** the default language instead.}%
\else
\language=\csname l@#1\endcsname
\fi
#2}}

\bibitem{Elmore2015}
{J. G. Elmore et~al.}, ``{Diagnostic Concordance Among Pathologists
  Interpreting Breast Biopsy Specimens},'' \emph{The Journal of the American
  Medical Association}, vol. 313, no.~11, pp. 1122--1132, 2015.

\bibitem{Filipczuk2013}
{P. Filipczuk et~al.}, ``{Computer-Aided Breast Cancer Diagnosis Based on the
  Analysis of Cytological Images of Fine Needle Biopsies},'' \emph{IEEE
  Transactions on Medical Imaging}, vol.~32, no.~12, pp. 2169--2178, 2013.

\bibitem{Veta2013}
{M. Veta et~al.}, ``{Automatic Nuclei Segmentation in H\&E Stained Breast
  Cancer Histopathology Images},'' \emph{Plos One}, vol.~8, no.~7, p. e70221,
  2013.

\bibitem{Ali2012}
{S. Ali and A. Madabhushi}, ``{An Integrated Region-, Boundary-, Shape-Based
  Active Contour for Multiple Object Overlap Resolution in Histological
  Imagery},'' \emph{IEEE Transactions on Medical Imaging}, vol.~31, no.~7, pp.
  1448--1460, 2012.

\bibitem{Kwak2015}
{J. T. Kwak et~al.}, ``{Nucleus Detection Using Gradient Orientation
  Information and Linear Least Squares Regression},'' in \emph{Proceedings of
  Medical Imaging: Digital Pathology}, vol. 9420, 2015, pp. 152--159.

\bibitem{Chang2018}
{C.-S. Chang et~al.}, ``{Cell Segmentation Algorithm Using Double Thresholding
  with Morphology-Based Techniques},'' in \emph{Proceedings of IEEE
  International Conference on Consumer Electronics, Taiwan}, 2018, pp. 1--5.

\bibitem{Song2018}
{J. Song, L. Xiao, and Z. Lian}, ``{Contour-Seed Pairs Learning-Based Framework
  for Simultaneously Detecting and Segmenting Various Overlapping Cells/Nuclei
  in Microscopy Images},'' \emph{IEEE Transactions on Image Processing},
  vol.~27, no.~12, pp. 5759--5774, 2018.

\bibitem{Song2019}
{J. Song et~al.}, ``{Multi-Layer Boosting Sparse Convolutional Model for
  Generalized Nuclear Segmentation from Histopathology Images},''
  \emph{Knowledge-Based Systems}, vol. 176, pp. 40--53, 2019.

\bibitem{Ronneberger2015}
O.~R. et~al., ``{U-Net: Convolutional Networks for Biomedical Image
  Segmentation},'' in \emph{Proceedings of Medical Image Computing and
  Computer-Assisted Intervention}, 2015, pp. 234--241.

\bibitem{Zhou2018}
{Z. Zhou et~al.}, ``{UNet++: A Nested U-Net Architecture for Medical Image
  Segmentation},'' in \emph{Proceedings of Deep Learning in Medical Image
  Analysis and Multimodal Learning for Clinical Decision Support}, 2018, pp.
  3--11.

\bibitem{Raza2019}
{S. E. A. Raza et~al.}, ``{Micro-Net: A Unified Model for Segmentation of
  Various Objects in Microscopy Images},'' \emph{Medical Image Analysis},
  vol.~52, pp. 160--173, 2019.

\bibitem{Preity2024}
{Preity, A. K. Bhandari and S. Shahnawazuddin}, ``{Soft Attention Mechanism
  Based Network to Extract Blood Vessels From Retinal Image Modality},''
  \emph{IEEE Transactions on Artificial Intelligence}, vol.~5, no.~7, pp.
  3408--3418, 2024.

\bibitem{He2017}
{K. He et~al.}, ``{Mask R-CNN},'' in \emph{Proceedings of IEEE International
  Conference on Computer Vision}, 2017, pp. 2961--2969.

\bibitem{Graham2019}
{S. Graham et~al.}, ``{Hover-Net: Simultaneous Segmentation and Classification
  of Nuclei in Multi-Tissue Histology Images},'' \emph{Medical Image Analysis},
  vol.~58, p. 101563, 2019.

\bibitem{Weigert2022}
{M. Weigert and U. Schmidt}, ``{Nuclei Instance Segmentation and Classification
  in Histopathology Images with Stardist},'' in \emph{Proceedings of IEEE
  International Symposium on Biomedical Imaging Challenges}, 2022, pp. 1--4.

\bibitem{Lijin2024}
{Lijin P. et~al.}, ``{Dual Encoder–Decoder Shifted Window-Based Transformer
  Network for Polyp Segmentation With Self-Learning Approach},'' \emph{IEEE
  Transactions on Artificial Intelligence}, vol.~5, no.~7, pp. 3456--3469,
  2024.

\bibitem{Qian2022}
{Z. Qian et~al.}, ``{Transformer Based Multiple Instance Learning for Weakly
  Supervised Histopathology Image Segmentation},'' in \emph{Proceedings of
  Medical Image Computing and Computer Assisted Intervention}, 2022, pp.
  160--170.

\bibitem{Lin2024}
{Y. Lin et~al.}, ``{BoNuS: Boundary Mining for Nuclei Segmentation with Partial
  Point Labels},'' \emph{IEEE Transactions on Medical Imaging}, vol.~43, no.~6,
  pp. 2137--2147, 2024.

\bibitem{Zhu2017}
{J. -Y. Zhu et~al.}, ``{Unpaired Image-to-Image Translation Using
  Cycle-Consistent Adversarial Networks},'' in \emph{Proceedings of IEEE
  International Conference on Computer Vision}, 2017, pp. 2242--2251.

\bibitem{Mahmood2020}
{F. Mahmood et~al.}, ``{Deep Adversarial Training for Multi-Organ Nuclei
  Segmentation in Histopathology Images},'' \emph{IEEE Transactions on Medical
  Imaging}, vol.~39, no.~11, pp. 3257--3267, 2020.

\bibitem{Liu2021}
{S. Liu et~al.}, ``{Unpaired Stain Transfer Using Pathology-Consistent
  Constrained Generative Adversarial Networks},'' \emph{IEEE Transactions on
  Medical Imaging}, vol.~40, no.~8, pp. 1977--1989, 2021.

\bibitem{Kwon2021}
{T. Kwon and J. C. Ye}, ``{Cycle-Free CycleGAN Using Invertible Generator for
  Unsupervised Low-Dose CT Denoising},'' \emph{IEEE Transactions on
  Computational Imaging}, vol.~7, pp. 1354--1368, 2021.

\bibitem{Villani2008}
{C. Villani}, \emph{{Optimal Transport: Old and New}}, ser. Grundlehren Der
  Mathematischen Wissenschaften.\hskip 1em plus 0.5em minus 0.4em\relax
  Springer Berlin Heidelberg, 2008.

\bibitem{Zhou2004}
{W. Zhou et~al.}, ``{Image Quality Assessment: From Error Visibility to
  Structural Similarity},'' \emph{IEEE Transactions on Image Processing},
  vol.~13, no.~4, pp. 600--612, 2004.

\bibitem{Otsu1979}
{N. Otsu}, ``{A Threshold Selection Method from Gray-Level Histograms},''
  \emph{IEEE Transactions on Systems, Man, and Cybernetics}, vol.~9, no.~1, pp.
  62--66, 1979.

\bibitem{Peyre2019}
{G. Peyr{\'e} and M. Cuturi}, \emph{{Computational Optimal Transport: With
  Applications to Data Science}}, ser. Foundations and Trends in Machine
  Learning.\hskip 1em plus 0.5em minus 0.4em\relax Now Publishers, 2019.

\bibitem{Cohn2013}
{D. L. Cohn}, \emph{{Measure Theory}}, ser. Birkh{\"a}user Advanced Texts
  Basler Lehrb{\"u}cher.\hskip 1em plus 0.5em minus 0.4em\relax Springer New
  York, 2013.

\bibitem{Dinh2017}
{L. Dinh and J. S. -Dickstein and S. Bengio}, ``{Density Estimation Using Real
  {NVP}},'' in \emph{Proceedings of International Conference on Learning
  Representations}, 2017.

\bibitem{Tomczak2021}
{J. M. Tomczak}, ``{General Invertible Transformations for Flow-based
  Generative Modeling},'' in \emph{Proceedings of International Conference on
  Machine Learning Workshop on Invertible Neural Networks, Normalizing Flows,
  and Explicit Likelihood Models}, 2021.

\bibitem{Kumar2017}
{N. Kumar et~al.}, ``{A Dataset and a Technique for Generalized Nuclear
  Segmentation for Computational Pathology},'' \emph{IEEE Transactions on
  Medical Imaging}, vol.~36, no.~7, pp. 1550--1560, 2017.

\bibitem{Graham2021}
{S. Graham et~al.}, ``{Conic: Colon nuclei identification and counting
  challenge 2022},'' \emph{arXiv preprint arXiv:2111.14485}, 2021.

\end{thebibliography}

\end{document}